\documentclass[11pt]{article}
\usepackage{amsmath, amsfonts, amsthm, amssymb}
\usepackage{hyperref}

\newcommand{\abs}[1]{{\lvert #1 \rvert}}
\newcommand{\norm}[1]{{\lVert #1 \rVert}}
\newcommand{\set}[1]{{\lbrace{#1}\rbrace}}

\newcommand{\braket}[2]{{\left\langle #1 \middle| #2 \right\rangle}}
\newcommand{\bra}[1]{{\left\langle #1 \right|}}
\newcommand{\ket}[1]{{\left| #1 \right\rangle}}
\newcommand{\tensor}{\otimes}
\newcommand{\Tensor}{\bigotimes}

\newcommand{\RR}{\mathbb{R}}
\newcommand{\CC}{\mathbb{C}}
\newcommand{\MM}{\mathbb{M}}
\newcommand{\eps}{\varepsilon}
\newcommand{\nya}{\tilde{n}}
\newcommand{\calM}{\mathcal{M}}
\newcommand{\calS}{\mathcal{S}}
\newcommand{\calB}{\mathcal{B}}
\newcommand{\Shat}{\hat{S}}
\newcommand{\MMya}{\widetilde{\mathbb{M}}}
\newcommand{\mya}{\tilde{m}}
\newcommand{\Zya}{\tilde{Z}}

\newcommand{\Cya}{\tilde{C}}
\newcommand{\Dya}{\tilde{D}}

\DeclareMathOperator{\EE}{\mathbb{E}}
\DeclareMathOperator{\Tr}{tr}
\DeclareMathOperator{\rank}{rank}
\DeclareMathOperator{\Var}{Var}

\theoremstyle{plain}
\newtheorem{definition}{Definition}[section]
\newtheorem{theorem}[definition]{Theorem}
\newtheorem{proposition}[definition]{Proposition}
\newtheorem{lemma}[definition]{Lemma}

\theoremstyle{definition}

\begin{document}

\title{Building one-time memories from isolated qubits}
\author{Yi-Kai Liu\\
Applied and Computational Mathematics Division\\
National Institute of Standards and Technology\\
Gaithersburg, MD, USA\\
yi-kai.liu@nist.gov}
\date{\today}
\maketitle

\abstract{\textit{One-time memories} (OTM's) are simple tamper-resistant cryptographic devices, which can be used to implement \textit{one-time programs}, a very general form of software protection and program obfuscation.  Here we investigate the possibility of building OTM's using quantum mechanical devices.  It is known that OTM's cannot exist in a fully-quantum world or in a fully-classical world.  Instead, we propose a new model based on \textit{isolated qubits} --- qubits that can only be accessed using \textit{local operations and classical communication} (LOCC).  This model combines a quantum resource (single-qubit measurements) with a classical restriction (on communication between qubits), and can be implemented using current technologies, such as nitrogen vacancy centers in diamond.  In this model, we construct OTM's that are information-theoretically secure against one-pass LOCC adversaries that use 2-outcome measurements.  



Our construction resembles Wiesner's old idea of quantum conjugate coding, implemented using random error-correcting codes; our proof of security uses entropy chaining to bound the supremum of a suitable empirical process.  In addition, we conjecture that our random codes can be replaced by some class of efficiently-decodable codes, to get computationally-efficient OTM's that are secure against computationally-bounded LOCC adversaries.

In addition, we construct data-hiding states, which allow an LOCC sender to encode an $(n-O(1))$-bit messsage into $n$ qubits, such that at most half of the message can be extracted by a one-pass LOCC receiver, but the whole message can be extracted by a general quantum receiver.}


\section{Introduction}


One-time memories (OTM's) are a simple type of tamper-resistant cryptographic hardware \cite{GKR}.  An OTM device behaves as follows:  one party (Alice) can write two messages $s,t\in\set{0,1}^k$ into the device, and then give the device to another party (Bob); after receiving the device, Bob can then choose to read either $s$ or $t$, but not both.  An OTM is far simpler than a general-purpose processor, but it can be used to implement sophisticated forms of secure computation, such as one-time programs\footnote{A one-time program is a package of hardware and software that is prepared by Alice and given to Bob.  It can compute a function $f$ (chosen by Alice when she prepares the package) on a single input $x$ provided by Bob (when he runs the package).  During its execution, the one-time program behaves like a black box, i.e., Bob learns nothing about its internal functioning.  After running once, the one-time program ``self-destructs,'' i.e., it stops functioning, and no more information can be extracted from it.} 
\cite{GKR,goyal,bellare} (and, more recently, quantum one-time programs \cite{broadbent}).
The remarkable fact about these constructions is that the OTM is the \textit{only} piece of hardware that has to be tamper-resistant; everything else consists of cryptographic software running on untrusted general-purpose processors.  

Intuitively, it seems much easier to build an OTM, rather than a general-purpose tamper-proof processor.  Indeed, there are many practical approaches to building such devices.  However, from a theoretical perspective, it would be nice if one could build \textit{provably-secure} OTM's based on some clear \textit{physical principle}, in the same way that one can build provably-secure encryption and signature schemes based on assumptions that certain problems are computationally intractable.  But this line of investigation runs into a number of obstacles.  OTM's cannot exist in a fully classical world, because information can always be copied without destroying it.  One might hope to build OTM's in a quantum world, where the no-cloning principle limits an adversary's ability to copy an unknown quantum state.  However, this is also impossible, because an OTM can be used to perform oblivious transfer with information-theoretic security, and there are strong no-go theorems for quantum oblivious transfer, quantum bit commitment, and many other kinds of two-party secure computation in a quantum world \cite{nogo1,nogo2,nogo3,nogo4}.  

One way around these no-go theorems is to try to construct protocols that are secure against a restricted class of quantum adversaries.  The adversaries in the no-go theorems (that break quantum bit commitment, oblivious transfer, etc.) seem to require the full power of a quantum computer, i.e., the ability to perform arbitrary quantum circuits with entangling gates.  However, a number of authors have shown protocols for bit commitment and other functionalities that are secure against adversaries who can only perform $k$-local measurements \cite{salvail}, or adversaries who only have bounded or noisy quantum storage \cite{bounded05, bounded06, bounded07, noisy07}.  

In this paper, we propose a new model of this type, called the \textit{isolated qubits} model.  This model allows the adversary to perform \textit{local operations and classical communication} (LOCC); intuitively, this is the class of operations that can generate classical correlations between the qubits, but not entanglement.  We then aim to construct OTM's that are secure in this model.  

The main challenge in this paper arises from the fact that OTM's are non-interactive:  after Alice gives the OTM to Bob, there is no further communication between them.  Thus, they cannot use standard cryptographic tools, such as privacy amplification, to give Bob an advantage over the adversary.  This makes OTM's very different from most other protocols for bit-commitment and oblivious transfer.  
\footnote{More precisely, the difference is that OTM's are asynchronous, in the sense that there is only one message from Alice to Bob.  In contrast, most other protocols (even those that only require one-way communication rather than two-way interaction \cite{tight-eur}) still use a sequence of two messages from Alice to Bob, in such a way that any dishonest action occurs before the second message, so that the second message can be used to perform privacy amplification.  A notable exception is the recent work \cite{all-but-one}, which considers a situation that is fully asynchronous, with a non-adaptive adversary.}

For our OTM's, we instead use an old idea called \textit{conjugate coding}, which is due to Wiesner \cite{wiesner}, and which works in this non-interactive setting.  Conjugate coding uses quantum states that are not entangled, hence they can be prepared by honest parties in our model.  Wiesner argued that conjugate coding \textit{is} secure against non-adaptive adversaries who can only perform single-qubit destructive measurements, but that it is \textit{not} secure against adversaries who can perform many-qubit entangled measurements.  However, apart from those two extreme cases, little seems to be known about the security of conjugate coding.  It is a natural question, then, whether conjugate coding \textit{is} secure against the more general class of LOCC adversaries (which may be adaptive, and may perform repeated weak measurements on the same qubit).  If the answer turned out to be yes, this would be a fairly realistic scenario in which conjugate coding provides useful security.

Unfortunately, proving good upper bounds on the power of LOCC adversaries is a long-standing open problem.  Previous results in this area include demonstrations of ``nonlocality without entanglement'' (NLWE) \cite{nlwe99} (see \cite{nlwe12} for a recent survey), and constructions of data-hiding states \cite{qdh02, eggeling02, hqd02, mpdhqi05}.  However, these results do not seem to apply to our OTM's.  We are only able to prove partial results on the security of our OTM's, but given that strong bounds of this type are quite rare, we believe this is significant progress.  

On a conceptual level, we show that information can be hidden from an LOCC adversary in a stronger and more sophisticated way, such that the adversary can retrieve one of the messages contained in the OTM, but not both.  This contrasts with previous work on NLWE and data-hiding states.  On a technical level, we prove security of our OTM's against ``1-pass'' LOCC adversaries (which may be adaptive, but are still restricted to destructive single-qubit measurements).  This extends Wiesner's original claim, though not all the way to general LOCC adversaries.  The techniques used to prove this result are quite nontrivial:  we construct our OTM's using random codes, and we prove security using entropy chaining (also called Dudley's inequality for suprema of empirical processes \cite{talagrand}, or ``using correlations to beat the union bound'').

\subsection{Isolated qubits}

In this paper we consider a model with \textit{isolated qubits}, where all parties are only allowed to perform local quantum operations (on each qubit) and classical communication (between qubits).  This class of operations is known as \textit{$n$-partite LOCC}, where $n$ is the number of qubits.  (See Section \ref{sec-rhinoceros} for details.)  We will construct an OTM that consists of $n$ isolated qubits.  When Alice prepares the device, she can perform $n$-partite LOCC operations on the qubits, and likewise, when Bob reads the device, he can perform $n$-partite LOCC operations on the qubits.  However, there is no communication or interaction between Alice and Bob, apart from the step where Alice gives the device (containing the $n$ qubits) to Bob.  

Note that this is a different scenario from most previous work on the power of LOCC operations \cite{nlwe99, qdh02}, where Alice and Bob share some \textit{bipartite} quantum system, and a ``local operation'' refers to an arbitrary operation on either Alice's subsystem or Bob's subsystem, and ``classical communication'' refers to communication between Alice and Bob.

Our model of isolated qubits is motivated by recent experimental work on nitrogen vacancy centers in diamond \cite{jelezko}.  Nitrogen vacancy (NV) centers can be used to implement single qubits that have relatively long coherence times (on the order of seconds or minutes), at room temperature in a solid-state material.  Individual NV centers can be read out and manipulated optically, but it is difficult to perform entangling operations on pairs of NV centers, due to variations in their emission spectra.  (Recent experiments have demonstrated entanglement between distant NV centers \cite{bernien2013}, but for our purposes there are natural ways of designing a device to prevent such entangling operations.)  NV centers have been studied in connection with quantum money \cite{pastawski}, and they are a plausible candidate to implement our model of isolated qubits.  

Isolated qubits are similar (but not directly comparable) to Salvail's $k$-local measurement model \cite{salvail}.  In Salvail's model, the adversary can perform entangled measurements on up to $k$ qubits, where $k$ is proportional to $n$; however, the adversary is only allowed to perform projective measurements, not repeated weak measurements (which are allowed in LOCC).  

We also argue that isolated qubits can exist in a world with quantum computers.  Isolated qubits are simply designed to satisfy different requirements than the qubits in a quantum computer.  More specifically, recall that in any quantum device, there is a tradeoff between two conflicting requirements:  first, protecting the device from unwanted interactions with the environment, such as noise and decoherence; and second, providing strong coherent interactions between the device and an external probe, in order to perform some useful task.  Isolated qubits represent one possible compromise between these requirements, namely strong protection from noise and decoherence, and only classical (not entangling) gates and measurements.  (Note that one cannot teleport information to or from an isolated qubit.)  In contrast, quantum computers and quantum memories must satisfy both of the above requirements, which is a different, possibly more difficult task.  (Indeed, the bounded / noisy storage model \cite{bounded05, noisy07} assumes that it is hard to build large, high-fidelity quantum memories.)  

In some sense, our isolated qubits model is complementary to the bounded / noisy storage model.  In the bounded / noisy storage model, the honest parties Alice and Bob use qubits that may allow entangled measurements, but cannot be stored for a long time.  In our model, Alice and Bob use qubits that can be stored for a long time, but do not allow entangled measurements.



\subsection{Data-hiding states}

Our first main result is a construction for data hiding states (see Section \ref{sec-triceratops}).  These states are simpler to analyze than our one-time memories, and they demonstrate the basic point that a sender can use LOCC operations to ``hide'' information from a LOCC receiver.  We consider a system of $n$ isolated qubits, and we construct a set of $2^{\nya}$ states, where $\nya := n-\Theta(1)$, by sampling independently at random from the set $\set{\ket{0},\ket{1},\ket{+},\ket{-}}^{\tensor n}$.  (Here, $\ket{+} := (\ket{0}+\ket{1})/\sqrt{2}$ and $\ket{-} := (\ket{0}-\ket{1})/\sqrt{2}$ are the Hadamard basis states.)  These states are all tensor products of single-qubit pure states, hence they can be prepared using only LOCC operations.  

First, we show that these states can be distinguished almost perfectly using an entangled quantum measurement (the ``pretty good measurement,'' see Section \ref{sec-triceratops-pgm}).  Then we consider ``one-pass'' LOCC measurement strategies, i.e., measurement strategies that measure each qubit at most once.  (For comparison, a general LOCC measurement strategy may perform many weak measurements on the same qubit.  Note that bounding the power of general LOCC measurement strategies is a difficult open problem.)  We show that a one-pass LOCC measurement strategy using 2-outcome measurements can extract at most $\approx n/2$ bits of information about which state was prepared\footnote{Formally, we upper-bound the Shannon mutual information.} (see Section \ref{sec-triceratops-locc2}).  Note that there exists a trivial LOCC measurement strategy that can extract $n/2$ bits of information, by measuring each qubit in the $\set{\ket{0},\ket{1}}$ basis, for instance; hence the above bound is tight.  In addition, we show that a one-pass LOCC measurement strategy using $q$-outcome measurements (for any constant $q$) can extract at most $\approx (0.7067)n$ bits of information (see Section \ref{sec-triceratops-loccq}).

The main point of this data-hiding result is to develop the proof techniques for our one-time memories, which will use a similar idea of sampling random states from the set $\set{\ket{0},\ket{1},\ket{+},\ket{-}}^{\tensor n}$, but will restrict access to the data in a more subtle way.  We use two proof techniques:  \textit{entropy chaining}, and a bound on the \textit{collision entropy}.  We will describe these techniques below.

In addition, our data-hiding states may also be of independent interest, as they differ from previous work in some significant ways.  On one hand, most previous constructions of data-hiding states \cite{qdh02, eggeling02, hqd02, mpdhqi05} are secure against a much stronger class of LOCC adversaries (with infinite LOCC rather than one-pass LOCC).  On the other hand, almost all of those constructions use entangled states, which cannot be realized in our isolated qubits model.  (An exception is \cite{eggeling02}, which uses separable Werner states.  This approach too is quite different from ours.)  

We remark that another line of work has focused on ``nonlocality without entanglement'' \cite{nlwe99}, where one considers a bipartite system, and one constructs sets of separable states that are orthogonal but cannot be perfectly distinguished using LOCC; see \cite{nlwe12} for a recent survey.  Finally, there are a number of elegant results about unambiguous state discrimination using multipartite LOCC, which are applicable when the number of states to be distinguished is relatively small \cite{walgate1, chen03, walgate2}.  

\subsection{Entropy chaining}

One of our proof techniques is ``entropy chaining,'' aka Dudley's inequality for empirical processes \cite{talagrand}.  This is similar to a union bound over the set of all one-pass adaptive LOCC measurement strategies, but it takes advantage of the positive correlations between the performance of strategies that are similar.  This approach gives a tight bound for adversaries that use 2-outcome measurements, but it performs poorly when applied to adversaries that use $q$-outcome measurements for large $q$ (see Section \ref{sec-triceratops-locc2}).  

The basic idea is as follows.  Let $E$ denote the random choices made in the construction of our data-hiding states.  Let the resulting collection of data-hiding states be denoted by $\ket{E(u)}$ (for all $u \in \set{0,1}^{\nya}$).  We imagine a game, where a referee chooses $u \in \set{0,1}^{\nya}$ uniformly at random, and prepares the state $\ket{E(u)}$, then an adversary performs some measurement strategy, and outputs a string of measurement outcomes $z$.  The adversary's goal is to maximize the mutual information $I(Z;U)$ (where $Z$ and $U$ are random variables containing the strings $z$ and $u$).  

As a first step, note that if we fix a particular adversary strategy, then with high probability over $E$, $I(Z;U) \approx n/2$.  To see this, write $I(Z;U) = H(Z) - H(Z|U)$; note that for 1-pass LOCC strategies using 2-outcome measurements, $H(Z) \leq n$; and note that $H(Z|U) = 2^{-\nya} \sum_u H(Z|U=u)$ is a sum of independent random variables with respect to $E$; hence by Hoeffding's inequality, with high probability over $E$, $H(Z|U)$ will be close to its expected value, which is roughly $n/2$.

We want to prove a much stronger statement, however.  We want to estimate the probability (over the random choice of $E$) that the \textit{best} LOCC strategy (chosen with knowledge of the states $\ket{E(u)}$) can extract more than $n/2$ bits of information about $U$.  To achieve this, we can try to use the union bound over all possible LOCC strategies.  However, note that adaptive LOCC strategies can be described by \textit{decision trees}, and in the case of 1-pass LOCC strategies using 2-outcome measurements, there are $2^{\Theta(2^n)}$ such decision trees.  Meanwhile, the quantity $H(Z|U)$ is a sum of only $2^{\nya}$ independent random variables, so the failure probability in Hoeffding's inequality is only exponentially small in $n$, not doubly-exponentially small in $n$.  Thus the union bound fails to give a useful result.

Entropy chaining fixes this problem by exploiting correlations among the different strategies --- the fact that two strategies that make similar measurements will produce similar results, and hence their failure probabilities do not add up in the worst-case fashion described by the union bound.  The term ``entropy chaining'' refers to the fact that one must use a sequence of these arguments, to capture both strong correlations between very similar strategies and weak correlations between less-similar strategies.  Each such argument involves covering the set of strategies with an $\eps$-net at a different resolution, which can be interpreted as bounding the entropy of the set.

\subsection{Bounding the collision entropy}

Our second proof technique involves calculating the collision entropy of the unknown message $U$, conditioned on every possible sequence of measurement outcomes.  This approach does not give a tight bound, but it works fairly well for all values of $q$ (see Section \ref{sec-triceratops-loccq}).  

Here we take a different perspective:  instead of considering LOCC measurement strategies (which correspond to decision trees), we consider measurement outcomes (which correspond to tensor products of single-qubit POVM elements).  That is, a measurement outcome is described by a POVM element of the form $M_A = \Tensor_{i\in A} M_i$, where $A \subset [n]$ is the set of qubits that were measured, and $M_i$ is a POVM element acting on qubit $i$.  (Measurement outcomes have this form when the adversary performs separable measurements, which include LOCC measurements as a special case; but this does not hold when the adversary performs entangled measurements.)

The basic idea is to fix some measurement outcome $M_A$, then lower-bound the collision entropy $H_2(U|M_A)$ (with high probability over $E$), and then use the union bound over all measurement outcomes $M_A$.  To lower-bound $H_2(U|M_A)$, we proceed as follows.  Essentially we want to upper-bound the collision probability 
\begin{equation}
\Pr[\text{col}|M_A] = \sum_u \Pr[U=u|M_A]^2 = 4^{-\nya} \Pr[M_A]^{-2} \sum_u \Pr[M_A|U=u]^2.
\end{equation}
To do this, we note that both $\Pr[M_A] = 2^{-\nya} \sum_u \Pr[M_A|U=u]$ and $\sum_u \Pr[M_A|U=u]^2$ are sums of independent random variables (with respect to $E$), and we use large deviation bounds.  Finally, to take the union bound over all $M_A$, we note that we only have to include those $M_A$ that are of tensor product form, hence the number of $M_A$ is exponential in $|A|$, rather than doubly exponential in $|A|$.  

\subsection{One-time memories}

We now describe our construction for one-time memories (see Section \ref{sec-stegosaurus-construction}).  We consider a system of $n$ isolated qubits, and we pick two random error-correcting codes, $C: \set{0,1}^k \rightarrow \set{0,1}^n$ and $D: \set{0,1}^k \rightarrow \set{0,1}^n$.  (That is, each codeword is chosen independently and uniformly at random in $\set{0,1}^n$.)  Given two messages $s$ and $t$ in $\set{0,1}^k$, we prepare each qubit $i$ (for $i=1,2,\ldots,n$) as follows.  Let $C(s)_i$ and $D(t)_i$ denote the $i$'th bit in the strings $C(s)$ and $D(t)$, respectively.  We prepare the $i$'th qubit in a pure state that has the following properties:  first, if the qubit is measured in the $\set{\ket{0},\ket{1}}$ basis, the outcome is more likely to be $\ket{0}$ if $C(s)_i = 0$, and $\ket{1}$ if $C(s)_i = 1$; and second, if the qubit is measured in the $\set{\ket{+},\ket{-}}$ basis, the outcome is more likely to be $\ket{+}$ if $D(t)_i = 0$, and $\ket{-}$ if $D(t)_i = 1$.  This is similar to Wiesner's idea of quantum conjugate coding \cite{wiesner}.  We refer to these states as one-time memory (OTM) states.  

It is straightforward to check that these OTM states can be prepared using only LOCC operations, and that an honest party can recover either $s$ or $t$ using only LOCC operations (see Section \ref{sec-stegosaurus-correct}).  

However, the security of these OTM states is somewhat problematic.  For instance, an LOCC adversary can always obtain partial information about both $s$ and $t$, by measuring some of the qubits in the $\set{\ket{0},\ket{1}}$ basis and some of the qubits in the $\set{\ket{+},\ket{-}}$ basis.  Also, these OTM states can ``leak'' extra information:  there is a one-pass LOCC strategy that can extract $n/2 \approx (1.2528)k$ bits of information about $s$ and $t$.  
\footnote{Let $\ket{\alpha_{C(s)_i D(t)_i}}$ be the state used to encode $C(s)_i$ and $D(t)_i$ into qubit $i$ (see Section 3 for the precise definition).  It turns out that $\ket{\alpha_{00}}$ and $\ket{\alpha_{11}}$ are orthogonal, and likewise, $\ket{\alpha_{01}}$ and $\ket{\alpha_{10}}$ are orthogonal.  So, a one-pass LOCC strategy that measures each qubit in the basis $\set{\ket{\alpha_{00}}, \ket{\alpha_{11}}}$ can extract $n/2 \approx (1.2528)k$ bits of information.}

To address this issue, we define a notion of a ``leaky OTM,'' which we believe is still strong enough to construct one-time programs (see Section \ref{sec-stegosaurus-leaky}).  Essentially, we conjecture that one-time programs can be built using Yao's garbled circuits \cite{GKR}, leaky OTM's, and a leak-resistant encryption scheme \cite{AGV}.  

We then present some evidence that our OTM states satisfy this notion of ``leaky security''.  Essentially, we prove that our OTM states satisfy a weaker notion of security, in which the smoothed min-entropy $H_\infty^\eps$ is replaced by the Shannon entropy $H$ (see Section \ref{sec-stegosaurus-construction}).  We believe it should be possible to strengthen this result to show ``leaky security.''  In particular, we note that certain parts of the proof already use the collision entropy $H_2$, which is stronger than $H$, and implies bounds on $H_\infty^\eps$.

Our technical result is that no 1-pass LOCC adversary using 2-outcome measurements can extract more than $\approx (1.9189)k < 2k$ bits of information about $(s,t)$ (see Section \ref{sec-stegosaurus-secure}).  (We believe that this constant factor can be improved.)  The proof uses a two-stage argument that applies both the collision entropy bound and entropy chaining; we will describe this below.

We remark that there is a subtle point involving the difference between 1-pass LOCC measurements and general LOCC measurements, when applied to our OTM's based on conjugate coding.  For our OTM's, there is a 1-pass LOCC measurement that can reconstruct $s$ (and there is a similar measurement for $t$).  Also, Winter's ``gentle measurement lemma'' \cite{gentle} implies that, if there is a \textit{nondestructive} measurement that reconstructs $s$, and there is a similar measurement for $t$, then there is a measurement that reconstructs both $s$ and $t$ simultaneously.  However, this does \textit{not} imply the existence of a 2-pass LOCC measurement that can reconstruct both $s$ and $t$ simultaneously.  

The reason is that, in order to reconstruct $s$ (or $t$) using 1-pass LOCC operations, the measurement must be destructive (i.e., one must perform a projective measurement on each qubit, obtain a string of classical measurement outcomes, and then run the classical decoding algorithm).  If one wants to reconstruct $s$ (or $t$) using a nondestructive measurement, one must use entangling operations (i.e., one must run the classical decoding algorithm on a superposition of many different inputs).  Thus the gentle measurement lemma cannot be applied to these particular 1-pass LOCC measurements, and it does not rule out the possibility that our OTM's are secure against 2-pass or general LOCC adversaries.

\subsection{Two-stage argument}

Let us denote our OTM states by $\ket{E(s,t)}$ (for $s,t\in\set{0,1}^k$).  These states resemble the data-hiding states studied previously, but there are some important differences.  First, there are fewer OTM states (there are $4^k$ states in dimension $2^n$, where $k \approx (0.3991)n$), hence the states are easier to distinguish.  Also, the OTM states are not constructed independently at random:  there are $4^k$ states $\ket{E(s,t)}$, but only $2\cdot 2^k$ independent random variables (consisting of the codewords $C(s)$ and $D(t)$).  As a result, there are positive correlations between states $\ket{E(s,t)}$ that have the same $s$ but different $t$.  

To deal with the correlations among the states $\ket{E(s,t)}$, we use large-deviation bounds for sums of locally dependent random variables \cite{janson, dubhashi}.  However, these large-deviation bounds are not as strong as the ones we had for data-hiding states, and so neither of our proof techniques (i.e., entropy chaining and collision entropy) gives a useful result by itself.  

To get around this difficulty, we combine the two techniques in sequence.  Let $S$ and $T$ be random variables containing the messages $s$ and $t$.  We use the collision entropy technique to analyze the first few steps taken by the adversary; this yields a lower-bound on $H_2(S,T|M_A)$, for any measurement outcome $M_A$ observed by the adversary thus far.  Then we use entropy chaining to prove bounds on the adversary's subsequent steps; this yields an upper-bound on $I(\tilde{Z};S,T|M_A)$, where $\tilde{Z}$ is the adversary's output from subsequent measurements.  It is necessary to apply the two techniques in this order, because the collision entropy technique yields an upper-bound on $\sum_{st} \Pr[S=s,T=t|M_A]^2$; this helps us to get stronger large-deviation bounds for the quantity 
\begin{equation}
H(\tilde{Z}|S,T,M_A) = \sum_{st} H(\tilde{Z}|S=s,T=t,M_A) \Pr[S=s,T=t|M_A], 
\end{equation}
which is crucial for entropy chaining.

\subsection{Outlook}

We think it is an interesting challenge to develop our OTM construction into a useful primitive for secure computation.  In this paper we have taken a first step, by constructing OTM's based on isolated qubits, and analyzing their security in a simple information-theoretic framework (e.g., using random codes in the OTM's, and describing the adversary's knowledge in terms of mutual information).  The next step is to make our OTM's efficient, and prove a stronger security guarantee that allows composition of OTM's to implement one-time programs.  

First, we conjecture that the random codes $C$ and $D$ can be replaced by some class of efficiently-decodable codes, to construct computationally-efficient one-time memories that are secure against computationally-bounded LOCC adversaries.  For comparison, note that the present construction, while not computationally efficient, also makes no assumptions about the adversary's computational power, i.e., it is secure against one-pass LOCC adversaries that have unbounded computational power.  

Second, we conjecture that our OTM's satisfy a particular notion of ``leaky security,'' which can be combined with leak-resistant encryption schemes \cite{AGV} to construct one-time programs.  This notion of ``leaky security'' uses the smoothed min-entropy to quantify the adversary's uncertainty about the messages $s$ and $t$.  Here we presented bounds that support this conjecture, using the Shannon entropy and the collision entropy.

Finally, it is an open problem to better understand the security of Wiesner's conjugate coding technique against general LOCC strategies (rather than the one-pass LOCC strategies considered here).  


\subsection{Notation}

For any integer $n \geq 1$, we define the set $[n] := \set{1,2,\ldots,n}$.  For any vector $v \in \CC^n$, we define the $\ell_2$ norm $\norm{v}_2 = (\sum_i \abs{v_i}^2)^{1/2}$.  

For any matrix $M \in \CC^{m\times n}$, with singular values $\lambda_1(M) \geq \lambda_2(M) \geq \cdots$, we define the operator norm $\norm{M} := \lambda_1(M)$ and the Frobenius norm $\norm{M}_F := (\sum_i \lambda_i(M)^2)^{1/2}$.  The notation $M \succeq 0$ means $M$ is positive semidefinite.  

An $\eps$-net $E$ (for a set $S$, with respect to some metric $d$) is a subset $E \subseteq S$ such that, for all $x\in S$, there exists some $x'\in E$, such that $d(x,x') \leq \eps$.  The covering number $N(S,d,\eps)$ is the minimum cardinality of any such $\eps$-net $E$.

Logarithms are denoted as follows:  $\ln(\cdot)$ is the natural logarithm, $\lg(\cdot)$ is the base-2 logarithm, and $\log(\cdot)$ is the logarithm when the base does not matter (because the log appears inside a big-O expression).

The Hamming distance between two binary strings $s,t\in\set{0,1}^n$ is denoted $d_H(s,t)$.

The $L_1$ or total variation distance between two random variables $X$ and $X'$ is denoted by $\Delta(X,X') = \sum_x \abs{\Pr[X=x] - \Pr[X'=x]}$.

The Shannon entropy of a random variable $X$ is denoted by $H(X) = -\sum_x \Pr[X=x] \lg\Pr[X=x]$, and the mutual information between random variables $X$ and $Y$ is denoted by $I(X;Y) = H(X) - H(X|Y)$.  (Note that $I$ without parentheses denotes the identity operator.  It will be clear from the context which one is meant.)  

The Renyi collision entropy of $X$ is denoted $H_2(X) = -\lg\bigl( \sum_x \Pr[X=x]^2 \bigr)$.  The min-entropy of $X$ is given by $H_\infty(X) = -\lg\bigl( \max_x \Pr[X=x] \bigr)$, and the smoothed min-entropy of $X$ is given by $H_\infty^\eps(X) = \max_{X' \;:\; \Delta(X,X')\leq\eps} H_\infty(X')$.


\section{Isolated qubits, and LOCC measurement strategies}
\label{sec-rhinoceros}

In this section we introduce the model of isolated qubits, and the class of LOCC measurement strategies.  Essentially, in a system of $n$ isolated qubits, the allowed operations are local (single-qubit) quantum operations, and classical communication between qubits.  These are $n$-party LOCC operations, where each party holds a single qubit.  

Any $n$-party LOCC measurement strategy can be described as a sequence of steps, which outputs a sequence of measurement outcomes, as follows:
\begin{verse}
Begin at step 1.\\
At step $a$, conditioned on the output of the previous steps $1,2,\ldots,a-1$:\\
\hspace{0.25in} Choose one of the parties, specified by $i\in\set{1,2,\ldots,n}$.\\
\hspace{0.25in} Choose some measurement $M$.
\footnote{Any measurement can be described by a set of measurement operators $K_1,K_2,\ldots$ which satisfy $\sum_j K_j^\dagger K_j = I$.  For a given state $\rho$, the measurement returns outcome $j$ with probability $\Tr(K_j\rho K_j^\dagger)$, and the post-measurement state (conditioned on observing $j$) is $K_j\rho K_j^\dagger / \Tr(K_j\rho K_j^\dagger)$.  Note that the measurement can also be described by a set of POVM elements $M_j = K_j^\dagger K_j$; the probability of observing outcome $j$ can then be written as $\Tr(M_j\rho)$.} \\
\hspace{0.25in} Perform the measurement $M$ on the $i$'th party's qubits; this yields some outcome $j$.\\
\hspace{0.25in} Output $j$, and proceed to step $a+1$.
\end{verse}

LOCC measurement strategies can use an unbounded number of steps, and can measure each qubit many times, for instance by using a sequence of weak measurements (which may be chosen adaptively).  Strategies using unbounded LOCC are difficult to analyze; in particular, it is a long-standing open problem to prove strong quantitative bounds on the amount of information returned by strategies using unbounded LOCC.  

Here we consider a restricted class of LOCC strategies:  those that measure each qubit \textit{at most once}.  We will refer to these as \textit{1-pass} LOCC strategies.  

Let us introduce some notation.  A 1-pass LOCC strategy consists of $n$ steps, labeled by $a \in [n]$ (where we define $[n] := \set{1,2,\ldots,n}$).  Suppose the strategy uses single-qubit measurements that have at most $q$ outcomes.  At step $a$, let $z_{<a} := (z_1,z_2,\ldots,z_{a-1}) \in [q]^{a-1}$ be the output of the previous steps; let $Q_a(z_{<a}) \in [n]$ be the choice of which qubit to measure next; let $M_a(z_{<a},\zeta) \in \CC^{2\times 2}$ (for all $\zeta \in [q]$) be the POVM elements corresponding to the choice of measurement in this step; and let $z_a$ be the actual measurement outcome that is obtained, so that $M_a(z_{\leq a})$ is the corresponding POVM element.  We can write the complete strategy as a POVM measurement on $n$ qubits, whose elements are given by 
\begin{equation}
M(z) := \Tensor_{a=1}^n M_a(z_{\leq a}), \quad \forall z\in[q]^n, 
\end{equation}
where $M_a(z_{\leq a})$ acts on the qubit indicated by $Q_a(z_{<a})$.  

\subsection{A state discrimination game}

Consider a collection of $n$-qubit quantum states $\ket{E(u)}$, indexed by $u \in \mathcal{U}$.  How well can an LOCC adversary distinguish among these states?  To make this question precise, one can define the following \textit{state discrimination game}:  first the referee chooses $u \in \mathcal{U}$ uniformly at random, then prepares the corresponding state $\ket{E(u)}$, and gives it to the adversary; then the adversary carries out some LOCC measurement strategy, and outputs some string $z\in[q]^n$.  We can measure the adversary's success in terms of the \textit{mutual information} $I(Z;U)$, where $U$ and $Z$ are the random variables describing the referee's choice and the adversary's output.

Note that, in the isolated qubits model, each party holds a single qubit (rather than a higher-dimensional quantum system).  We can make use of this fact, to further simplify the set of possible LOCC strategies.  

\begin{lemma}\label{lem-locc-rank1}
Let $\calM$ be any 1-pass LOCC strategy in the isolated qubits model, which uses $q$-outcome measurements and returns output $Z$.  Then there exists $\calM'$, a 1-pass LOCC strategy in the isolated qubits model, which uses $q$-outcome measurements and returns output $Z'$, and has the following additional properties:  
\begin{enumerate}
\item $I(Z';U) \geq I(Z;U)$ (when playing the state discrimination game shown above).
\item In every measurement performed by $\calM'$, the POVM elements all have rank 1.
\end{enumerate}
\end{lemma}

\noindent
Proof:  See Appendix \ref{app-locc-epsnet}.



\subsection{Discretization of LOCC strategies}

Let $\calS$ be the set of all single-qubit measurements with $q$ outcomes, where every POVM element has rank 1:  
\begin{equation}
\calS = \set{(M_1,\ldots,M_q) \;|\; M_i \in \CC^{2\times 2}, \; M_i \succeq 0, \; \sum_{i=1}^q M_i = I, \; \rank(M_i) = 1}.
\end{equation}
This is a continuous set.  In our proofs, we would like to approximate it by a finite $\eps$-net $L$, with respect to some appropriate metric $t$.  It will be convenient to define $t$ as follows:
\begin{equation}
t(M,\tilde{M}) := \max_{i\in[q]} \norm{M_i - \tilde{M}_i}.
\end{equation}
(Here $\norm{\cdot}$ denotes the operator norm.)  The following two lemmas bound the size of the $\eps$-net $L$, first in the special case where $q=2$ (for which we have a better bound), and then in the general case where $q \geq 2$.

\begin{lemma}\label{lem-epsnet-2}
Let $q = 2$.  For any $0 < \eps \leq 1$, there exists an $\eps$-net $L$ for $\calS$, with respect to the metric $t$, that has cardinality $|L| \leq C/\eps^2$ (where $C$ is some numerical constant).  Equivalently, we have $N(\calS,t,\eps) \leq C/\eps^2$.
\end{lemma}

\noindent
Proof:  See Appendix \ref{app-locc-epsnet}.

\begin{lemma}\label{lem-epsnet-q}
Let $q \geq 2$.  For any $0 < \eps \leq 1$, there exists an $\eps$-net $L$ for $\calS$, with respect to the metric $t$, that has cardinality $|L| \leq (C/\eps)^{3q}$ (where $C$ is some numerical constant).  Equivalently, we have $N(\calS,t,\eps) \leq (C/\eps)^{3q}$.
\end{lemma}

\noindent
Proof:  See Appendix \ref{app-locc-epsnet}.

{\vskip 11pt}

We now bound the effect of this discretization when applied to a complete LOCC strategy.  Essentially, if we choose $\eps \leq O(1/qn)$, then the discretization has a negligible effect on the amount of information returned by the strategy.

\begin{lemma}\label{lem-locc-epsnet}
Let $\calM$ be any 1-pass LOCC strategy in the isolated qubits model, which uses $q$-outcome measurements, where all POVM elements have rank 1, and which has output $Z$.  Fix some $0 < \eps \leq 1/(qne)$, and let $L$ be the $\eps$-net for $\calS$ defined above.  Let $\calM'$ be the strategy that is obtained by duplicating the strategy $\calM$, and replacing each measurement $M \in \calS$ with the best approximating measurement $\tilde{M} \in L$.  Let $Z'$ be the output of the strategy $\calM'$.  Then 
\begin{equation}
\abs{I(Z';U) - I(Z;U)} \leq 2qn^2\eps + 2\eta(qn\eps), 
\end{equation}
where $\eta(x) := -x\lg x$.
\end{lemma}

\noindent
Proof:  See Appendix \ref{app-locc-epsnet}.


\section{Data-hiding states}
\label{sec-triceratops}

Consider a system of $n$ qubits.  We will construct a set $B$ of $2^{\nya}$ quantum states, with $\nya \geq n - O(1)$, that has the following properties:  
\begin{enumerate}
\item The states are pure and unentangled (i.e., they are tensor products of pure single-qubit states).
\item There exists an entangled quantum measurement that distinguishes these states almost perfectly.  In particular, given a state chosen uniformly at random from $B$, this measurement recovers nearly $\nya$ bits of information about the identity of the state.
\item No $n$-party LOCC measurement strategy can distinguish these states very well.  In particular, given a state chosen uniformly at random from $B$, no $n$-party LOCC measurement strategy using 2-outcome measurements can recover more than about $n/2$ bits of information about the identity of the state.  Similar bounds hold for $n$-party LOCC measurement strategies using $q$-outcome measurements, for constant $q$.
\end{enumerate}

We construct the set of states $B$ as follows.  Set $\nya = n - \Theta(1)$.  Briefly, $B$ is a set of $2^{\nya}$ states chosen independently and uniformly at random from the set $\set{\ket{0},\ket{1},\ket{+},\ket{-}}^{\tensor n}$.  To state this more explicitly, we define the following single-qubit states: 
\begin{equation}
\ket{\alpha_{00}} := \ket{0}, \quad
\ket{\alpha_{11}} := \ket{1}, \quad
\ket{\alpha_{01}} := \ket{+} = \tfrac{1}{\sqrt{2}} (\ket{0}+\ket{1}), \quad
\ket{\alpha_{10}} := \ket{-} = \tfrac{1}{\sqrt{2}} (\ket{0}-\ket{1}).
\end{equation}
Choose a random mapping $E: \set{0,1}^{\nya} \rightarrow \set{00,01,10,11}^n$, i.e., for each $u \in \set{0,1}^{\nya}$, assign $E(u)$ a value chosen independently and uniformly at random in $\set{00,01,10,11}^n$.  Also, for $a=1,2,\ldots,n$, let $E(u)_a \in \set{00,01,10,11}$ denote the $a$'th entry in the string $E(u)$.  Then let $B$ be the set of states $\ket{E(u)}$ defined as follows:  
\begin{equation}
\ket{E(u)} := \Tensor_{a=1}^n \ket{\alpha_{E(u)_a}}, \quad \forall u \in \set{0,1}^{\nya}.
\end{equation}

We will consider the following state discrimination problem, which we describe as a game between a referee and a distinguisher.  First, the referee chooses a random string $u$ in $\set{0,1}^{\nya}$, and prepares the state $\ket{E(u)}$.  Given this state, the distinguisher performs some measurement, and outputs a string $z$ (over some alphabet).  The goal of the distinguisher is to maximize the \textit{mutual information} $I(Z;U)$, where $U$ and $Z$ are the random variables representing the referee's choice of the state and the distinguisher's output.


\subsection{The pretty good measurement}
\label{sec-triceratops-pgm}

In this section we show that the states $\ket{E(u)}$ can be distinguished almost perfectly by a measurement that uses entanglement among the $n$ qubits.  In particular, we will consider the ``pretty good measurement'' \cite{pgm}, which is defined as follows.  Let $\rho$ be the mixed state $\rho := 2^{-\nya} \sum_{u\in\set{0,1}^{\nya}} \ket{E(u)}\bra{E(u)}$.  Then the ``pretty good measurement'' is given by the following set of POVM elements:
\begin{equation}
M_{\text{PGM}} := \set{\ket{M(z)}\bra{M(z)}, \; z\in\set{0,1}^{\nya}}, \text{ where } 
\ket{M(z)} := 2^{-\nya/2}\rho^{-1/2}\ket{E(z)}.
\end{equation}
(If $\rho$ is not full-rank, then $\rho^{-1/2}$ is defined on the support of $\rho$.)

We will show that, with high probability over the choice of the states $\ket{E(u)}$, the pretty good measurement works well.  In particular, let $Z$ be the output of the pretty good measurement; we will show that $Z=U$ with probability close to 1, and the mutual information $I(Z;U)$ is close to $\nya$.  

\begin{lemma}
Let $C \geq 1$.  With probability $\geq 1-\frac{1}{C}$ (over the choice of $E$), we have 
\begin{equation}\label{eq-gecko-0}
\Pr[Z=U] \geq 1 - 2\sqrt{C} \cdot 2^{(\nya-n)/2}.
\end{equation}
In particular, for any $\eps>0$, suppose that $\nya$ satisfies $\nya \leq n-\lg(C/\eps^2)-2$.  Then equation (\ref{eq-gecko-0}) implies that 
\begin{equation}
\Pr[Z=U] \geq 1-\eps.
\end{equation}
\end{lemma}

\noindent
Proof:  First, we will give a lower-bound for $\Pr[Z=U]$ in terms of the eigenvalues of the Gram matrix of the states $\ket{E(u)}$, using an argument due to Montanaro \cite{montanaro}.  We write 
\begin{equation}
\Pr[Z=U] = 2^{-\nya} \sum_{u \in \set{0,1}^{\nya}} \abs{\braket{M(u)}{E(u)}}^2
 = 4^{-\nya} \sum_{u \in \set{0,1}^{\nya}} \abs{\bra{E(u)} \rho^{-1/2} \ket{E(u)}}^2.
\end{equation}
We define the matrix $P \in \CC^{2^{\nya} \times 2^{\nya}}$, whose entries are $P_{uv} = \bra{E(u)} \rho^{-1/2} \ket{E(v)}$.  In addition, we define the Gram matrix $G \in \CC^{2^{\nya} \times 2^{\nya}}$, whose entries are $G_{uv} = \braket{E(u)}{E(v)}$.  It is easy to see that both $P$ and $G$ are positive semidefinite, and that $P^2 = 2^{\nya} G$; hence we can write $P = 2^{\nya/2} \sqrt{G}$.  So we have 
\begin{equation}
\Pr[Z=U] = 4^{-\nya} \sum_u \abs{P_{uu}}^2
 = 2^{-\nya} \sum_u ((\sqrt{G})_{uu})^2.
\end{equation}
We can lower-bound this as follows, using the convexity of the square function, and letting $\lambda_u(G)$ denote the eigenvalues of $G$:
\begin{equation}
\Pr[Z=U] \geq \Bigl( 2^{-\nya} \sum_u (\sqrt{G})_{uu} \Bigr)^2
 = \Bigl( 2^{-\nya} \Tr \sqrt{G} \Bigr)^2
 = \Bigl( 2^{-\nya} \sum_u \sqrt{\lambda_u(G)} \Bigr)^2.
\end{equation}

Next, define 
\begin{equation}
\nu_u(G) := \begin{cases}
\lambda_u(G) &\text{if } 0 \leq \lambda_u(G) \leq 1 \\
1 &\text{if } \lambda_u(G) \geq 1
\end{cases}
\end{equation}
and observe that $\sqrt{\lambda_u(G)} \geq \nu_u(G)$.  So we have 
\begin{equation}
\begin{split}
\Pr[Z=U] &\geq \Bigl( 2^{-\nya} \sum_u \nu_u(G) \Bigr)^2
 = \Bigl( 1 - 2^{-\nya} \sum_u (1-\nu_u(G)) \Bigr)^2 \\
 &\geq 1 - 2\cdot 2^{-\nya} \sum_u (1-\nu_u(G))
 \geq 1 - 2\cdot 2^{-\nya} 2^{\nya/2} \norm{\vec{1} - \vec{\nu}(G)}_2.
\end{split}
\end{equation}
Also note that $\abs{1-\nu_u(G)} \leq \abs{1-\lambda_u(G)}$, hence $\norm{\vec{1} - \vec{\nu}(G)}_2 \leq \norm{\vec{1} - \vec{\lambda}(G)}_2 = \norm{G-I}_F$.  So we have 
\begin{equation}\label{eq-gecko}
\Pr[Z=U] \geq 1 - 2\cdot 2^{-\nya/2} \norm{G-I}_F.
\end{equation}

Finally, we will use Markov's inequality to show that, with high probability (over the choice of $E$), $\norm{G-I}_F$ is not too large.  We write:
\begin{equation}
\EE_E[\norm{G-I}_F^2] = \EE_E\Bigl[ \sum_{u\neq v \in \set{0,1}^{\nya}} \abs{G_{uv}}^2 \Bigr], 
\end{equation}
\begin{equation}
\EE_E[\abs{G_{uv}}^2] = \EE_E\Bigl[ \prod_{a=1}^n \abs{\braket{\alpha_{E(u)_a}}{\alpha_{E(v)_a}}}^2 \Bigr]
 = \prod_{a=1}^n (\tfrac{1}{4}(1+0+\tfrac{1}{2}+\tfrac{1}{2})) = 2^{-n} \quad (\forall u\neq v), 
\end{equation}
\begin{equation}
\EE_E[\norm{G-I}_F^2] = 2^{\nya} (2^{\nya}-1) 2^{-n} < 4^{\nya} 2^{-n}.
\end{equation}
Hence, by Markov's inequality, for any $C\geq 1$, 
\begin{equation}
\Pr_E[\norm{G-I}_F^2 \geq C\cdot 4^{\nya} 2^{-n}] \leq \tfrac{1}{C}.
\end{equation}
That is, with probability $\geq 1-\tfrac{1}{C}$ (over the choice of $E$), we have $\norm{G-I}_F \leq \sqrt{C}\cdot 2^{\nya} 2^{-n/2}$.  Combining this with equation (\ref{eq-gecko}) completes the proof.  $\square$

{\vskip 11pt}

We have just shown a lower-bound on the success probability $\Pr[Z=U]$; this now implies a lower-bound on the mutual information $I(Z;U)$.  

\begin{lemma}\label{lem-ostrich}
Suppose that $\Pr[Z=U] \geq 1-\eps$, and $\eps$ is sufficiently small that $2\sqrt{\eps} + 2^{-\nya} \leq 1/e$.  Then $I(Z;U) \geq (1-5\sqrt{\eps}) \nya - \eta(2\sqrt{\eps})$, where $\eta(x) := -x \lg x$.
\end{lemma}

\noindent
Proof:  See Appendix \ref{app-ostrich}.


\subsection{1-pass LOCC strategies with 2-outcome measurements}
\label{sec-triceratops-locc2}

In this section, we prove that the states $\ket{E(u)}$ ($u \in \set{0,1}^{\nya}$) cannot be fully distinguished by any 1-pass LOCC strategy that uses 2-outcome measurements.  In particular, we show that any such strategy cannot extract more than about $n/2$ bits of information about $U$.  This claim holds with high probability over the randomized construction of the states $\ket{E(u)}$; more precisely, the claim holds with high probability over the choice of the map $E: \set{0,1}^{\nya} \rightarrow \set{00,01,10,11}^n$, which we view as a random variable.

The proof uses an entropy chaining argument, which is stated in Lemma \ref{lem-dudley}.  This is essentially Dudley's inequality for bounding the supremum of an empirical process with Gaussian decaying correlations \cite{talagrand}, with some minor technical modifications (in particular, the result is stated as a tail bound for the supremum, rather than a bound on the expected supremum); the proof is given in Appendix \ref{app-entropy-chaining}.

\begin{theorem}\label{thm-willow}
Let $\MM$ be the set of all 1-pass LOCC strategies in the isolated qubits model using 2-outcome measurements.  Let $t_0 > 0$ and $u \geq 1$.  Then, with probability $\geq 1 - \exp(-2t_0^2) - 2\cdot 2^{-u^2}$ (over the choice of $E$), the following statement holds:
\begin{equation}
\forall \calM \in \MM, \; I(Z;U) \leq (0.54)n + O(1) + t_0\nya 2^{-\nya/2} + u\cdot O(\sqrt{\log n}), 
\end{equation}
where $Z$ denotes the output of the strategy $\calM$.
\end{theorem}

\noindent
Proof:  First, let $L$ be an $\eps$-net for the set of single-qubit measurements with 2 outcomes where all POVM elements have rank 1, as described in Lemma \ref{lem-epsnet-2}; and set $\eps = 1/(100n)$.  Let $\MM'$ be the set of all strategies that use measurements chosen from the set $L$.  By Lemmas \ref{lem-locc-rank1} and \ref{lem-locc-epsnet}, any strategy in $\MM$ can be approximated by one in $\MM'$.

For any strategy $\calM$, let $Z$ denote its output.  Note that $Z$ takes values in $\set{0,1}^n$, and we can split its output into two pieces, $Z = (Z_{1,\ldots,\nya}, Z_{\nya+1,\ldots,n})$.  So we can write
\begin{equation}\label{eq-sparrow-1}
I(Z;U) = H(Z) - H(Z|U) \leq n - H(Z_{1,\ldots,\nya}|U).
\end{equation}
We want to show that $H(Z_{1,\ldots,\nya}|U)$ is not too small.

Let $\MM''$ be the set of all strategies with $\nya$ steps, whose behavior matches the first $\nya$ steps of some strategy in $\MM'$.  For any $\calM \in \MM''$, we now define 
\begin{equation}
Q_\calM := H(Z_{1,\ldots,\nya}|U), 
\end{equation}
which is a random variable depending on $E$.  Let $\mu_\calM := \EE_E Q_\calM$; we will prove a lower bound for $\mu_\calM$ below.  We will then use entropy chaining (Lemma \ref{lem-dudley}) to lower-bound the quantity
\begin{equation}
\inf_{\calM \in \MM''} (Q_\calM - \mu_\calM).
\end{equation}

First, we evaluate $\mu_\calM$:
\begin{equation}
\mu_\calM = \EE_E \bigl[ H(Z_{1,\ldots,\nya}|U) \bigr]
 = 2^{-\nya} \sum_{u\in\set{0,1}^{\nya}} \sum_{z\in\set{0,1}^{\nya}}
 \EE_E \bigl[ -\Pr(z|u) \lg \Pr(z|u) \bigr],
\end{equation}
where for convenience we wrote $\Pr(z|u)$ in place of $\Pr[Z_{1,\ldots,\nya}=z|U=u]$.  Consider any fixed $u, z \in \set{0,1}^{\nya}$.  Recall that $\ket{E(u)}$ is chosen uniformly at random in $A^{\tensor n}$, where $A := \set{\ket{\alpha_{00}}, \ket{\alpha_{01}}, \ket{\alpha_{10}}, \ket{\alpha_{11}}}$.  So we have 
\begin{equation}
\EE_E \bigl[ -\Pr(z|u) \lg \Pr(z|u) \bigr]
 = 4^{-\nya} \sum_{\ket{\psi} \in A^{\tensor \nya}} -\bra{\psi}M(z)\ket{\psi} \lg \bra{\psi}M(z)\ket{\psi}.
\end{equation}
Furthermore, we know that $\bra{\psi}M(z)\ket{\psi} = \prod_{a=1}^{\nya} \bra{\psi_{Q_a(z_{<a})}} M_a(z_{\leq a}) \ket{\psi_{Q_a(z_{<a})}}$, where we used the notation introduced in Section \ref{sec-rhinoceros}.  Hence we can write
\begin{equation}
\begin{split}
\EE_E \bigl[ & -\Pr(z|u) \lg \Pr(z|u) \bigr] \\
 &= \sum_{a=1}^{\nya} 4^{-\nya} \sum_{\ket{\psi} \in A^{\tensor \nya}} -\bra{\psi}M(z)\ket{\psi} \lg \bra{\psi_{Q_a(z_{<a})}} M_a(z_{\leq a}) \ket{\psi_{Q_a(z_{<a})}} \\
 &= \sum_{a=1}^{\nya} 4^{-\nya} \Bigl[ \sum_{\psi_a \in A} -\bra{\psi_a} M_a(z_{\leq a}) \ket{\psi_a} \lg \bra{\psi_a} M_a(z_{\leq a}) \ket{\psi_a} \Bigr] 
\prod_{b\neq a} \Bigl[ \sum_{\psi_b \in A} \bra{\psi_b} M_b(z_{\leq b}) \ket{\psi_b} \Bigr] 
\end{split}
\end{equation}
Recall that we are considering single-qubit measurements with 2 outcomes, where each outcome corresponds to a rank-1 POVM element.  Hence each $M_b(z_{\leq b})$ is a rank-1 projector, i.e., it can be viewed as a density matrix of a quantum state.  Hence we can write 
\begin{equation}
\sum_{\psi_b \in A} \bra{\psi_b} M_b(z_{\leq b}) \ket{\psi_b} = 2\Tr M_b(z_{\leq b}) = 2.
\end{equation}
Also, suppose we let $R_0$ be the result of measuring the state $M_a(z_{\leq a})$ in the orthonormal basis $\set{\ket{\alpha_{00}}, \ket{\alpha_{11}}}$, and we let $R_1$ be the result of measuring the same state in the orthonormal basis $\set{\ket{\alpha_{01}}, \ket{\alpha_{10}}}$.  Then we can write 
\begin{equation}
\sum_{\psi_a \in A} -\bra{\psi_a} M_a(z_{\leq a}) \ket{\psi_a} \lg \bra{\psi_a} M_a(z_{\leq a}) \ket{\psi_a}
 = H(R_0) + H(R_1) \geq 1, 
\end{equation}
using an entropic uncertainty relation of Maassen and Uffink \cite{maassen-uffink, wehner-winter}.  Substituting into the previous equations, we get 
\begin{equation}
\EE_E \bigl[ -\Pr(z|u) \lg \Pr(z|u) \bigr] \geq 2^{-\nya-1} \cdot \nya,
\end{equation}
\begin{equation}
\mu_\calM \geq \nya/2.
\end{equation}

We now show several technical facts which are needed in order to apply the entropy chaining argument (Lemma \ref{lem-dudley}).  First, fix some particular strategy $\calM_0 \in \MM''$.  We will show that $Q_{\calM_0}$ is tightly concentrated around its mean.  Observe that $Q_{\calM_0} = 2^{-\nya} \sum_{u\in\set{0,1}^{\nya}} H(Z_{1,\ldots,\nya}|U=u)$ is a sum of $2^{\nya}$ independent random variables, since the strings $E(u)$ for different $u$ are chosen independently at random.  Using Hoeffding's inequality, we get that 
\begin{equation}
\Pr_E[Q_{\calM_0} < \mu_{\calM_0} - t] \leq \exp\Bigl( -\frac{2t^2}{2^{\nya} (n/2^{\nya})^2} \Bigr)
= \exp\Bigl( -\frac{2t^2 \cdot 2^{\nya}}{\nya^2} \Bigr) \quad (\forall t>0),
\end{equation}
or equivalently
\begin{equation}
\Pr_E\Bigl[Q_{\calM_0} < \mu_{\calM_0} - \frac{t_0 \nya}{2^{\nya/2}}\Bigr] \leq \exp(-2t_0^2), \quad (\forall t_0>0).
\end{equation}

Next, we show that when two strategies $\calM$ and $\calM'$ are ``similar,'' the random variables $Q_\calM$ and $Q_{\calM'}$ are positively correlated.  In particular, suppose that $\calM$ and $\calM'$ behave identically for the first $\ell$ steps.  Let $Z_{1,\ldots,\nya}$ and $Z'_{1,\ldots,\nya}$ be the output of these two strategies; then $(Z_{1,\ldots,\ell},U)$ and $(Z'_{1,\ldots,\ell},U)$ have the same distribution.  So we can write
\begin{equation}
\begin{split}
Q_\calM - Q_{\calM'}
 &= H(Z_{\ell+1,\ldots,\nya}|Z_{1,\ldots,\ell},U) - H(Z'_{\ell+1,\ldots,\nya}|Z'_{1,\ldots,\ell},U) \\
 &= 2^{-\nya} \sum_{u\in\set{0,1}^{\nya}} H(Z_{\ell+1,\ldots,\nya}|Z_{1,\ldots,\ell},U=u) - H(Z'_{\ell+1,\ldots,\nya}|Z'_{1,\ldots,\ell},U=u), 
\end{split}
\end{equation}
which again is a sum of $2^{\nya}$ independent random variables.  By Hoeffding's inequality, 
\begin{equation}
\Pr_E[Q_\calM - Q_{\calM'} - \mu_\calM + \mu_{\calM'} \geq t]
 \leq \exp\Bigl( -\frac{2t^2}{2^{\nya} (2(\nya-\ell)/2^{\nya})^2} \Bigr)
 = \exp\Bigl( -\frac{t^2 \cdot 2^{\nya}}{2 (\nya-\ell)^2} \Bigr).
\end{equation}
We can rewrite this bound in terms of a metric $d$ that measures the ``distance'' between strategies.  We define $d$ as follows:
\footnote{Note that $d$ is indeed a metric:  It is easy to see that $d(\calM,\calM') \geq 0$, with equality iff $\calM = \calM'$.  Also, clearly $d(\calM,\calM') = d(\calM',\calM)$.  It remains to show that $d(\calM,\calM'') \leq d(\calM,\calM') + d(\calM',\calM'')$.  We consider two cases.  On one hand, if $\calM'$ satisfies $\ell(\calM,\calM') \leq \ell(\calM,\calM'')$ or $\ell(\calM',\calM'') \leq \ell(\calM,\calM'')$, then the claim follows immediately.  On the other hand, if $\calM'$ satisfies $\ell(\calM,\calM') > \ell(\calM,\calM'')$ and $\ell(\calM',\calM'') > \ell(\calM,\calM'')$, then this is impossible, since $\calM$ and $\calM''$ do not agree at step $\ell(\calM,\calM'')+1$; hence this case cannot occur.}
\begin{equation}
d(\calM,\calM') := \sqrt{2} \cdot 2^{-\nya/2} (\nya-\ell(\calM,\calM')), \text{ where}
\end{equation}
\begin{equation}
\ell(\calM,\calM') := \max \set{\ell \;|\; \text{$0\leq\ell\leq\nya$, and $\calM$ and $\calM'$ behave identically on steps $1,2,\ldots,\ell$}}.
\end{equation}
We then have 
\begin{equation}
\Pr_E[Q_\calM - Q_{\calM'} - \mu_\calM + \mu_{\calM'} \geq t]
 \leq \exp\Bigl( -\frac{t^2}{d(\calM,\calM')^2} \Bigr).
\end{equation}

Next, we bound the covering numbers of $\MM''$ with respect to the metric $d$.  We use a simple bound:
\begin{equation}
N(\MM'', d, \sqrt{2} \cdot 2^{-\nya/2}\eps)
 \leq \begin{cases}
(n\abs{L})^{2^{\lceil\nya-\eps\rceil}} &\text{if } \eps\leq\nya \\
1 &\text{if } \eps>\nya.
\end{cases}
\end{equation}
(This bound simply counts the number of possible strategies with $\lceil\nya-\eps\rceil$ steps.  Each such strategy is described by a binary tree of depth $\lceil\nya-\eps\rceil$, and at every node there is a choice of which of the $n$ qubits to measure next, and which of the measurements in the set $L$ to perform.)  We now bound the integral appearing in Lemma \ref{lem-dudley} (``Dudley's entropy integral''):  (here $C_0$ is a numerical constant)
\begin{equation}
\begin{split}
S &\leq C_0 \int_0^\infty \sqrt{\log N(\MM'',d,\eps)} d\eps \\
 &= C_0 \int_0^\infty \sqrt{\log N(\MM'',d,\sqrt{2} \cdot 2^{-\nya/2} \eps)} \cdot \sqrt{2} \cdot 2^{-\nya/2} d\eps \\
 &\leq C_0 \int_0^{\nya} \sqrt{2^{\nya-\eps+1} \log(n\abs{L})} \cdot \sqrt{2} \cdot 2^{-\nya/2} d\eps \\
 &= C_0 \int_0^{\nya} 2^{-\eps/2} d\eps \cdot 2\sqrt{\log(n\abs{L})} \\
 &= C_0 (\tfrac{2}{\ln 2}) (1-2^{-\nya/2}) \cdot 2\sqrt{\log(n\abs{L})} \\
 &< C_0 (\tfrac{2}{\ln 2}) \cdot 2\sqrt{\log(n\abs{L})}.
\end{split}
\end{equation}
Recall that $\abs{L} \leq O(1/\eps^2) = O(n^2)$.  Hence we have:  
\begin{equation}
S \leq O(\sqrt{\log n}).
\end{equation}

Finally, using Lemma \ref{lem-dudley}, we have that:  for all $t_0>0$ and $u \geq 1$, with probability $\geq 1 - \exp(-2t_0^2) - 2\cdot 2^{-u^2}$, the following holds:
\begin{equation}
\forall \calM \in \MM'', \; Q_{\calM} - \mu_{\calM} > - \frac{t_0\nya}{2^{\nya/2}} - uS.
\end{equation}
This implies that
\begin{equation}
\forall \calM \in \MM'', \; Q_{\calM} > \frac{\nya}{2} - \frac{t_0\nya}{2^{\nya/2}} - u\cdot O(\sqrt{\log n}).
\end{equation}
Plugging into equation (\ref{eq-sparrow-1}), we get that
\begin{equation}
\forall \calM \in \MM', \; I(Z;U) < n - \frac{\nya}{2} + \frac{t_0\nya}{2^{\nya/2}} + u\cdot O(\sqrt{\log n}).
\end{equation}
Finally, using Lemmas \ref{lem-locc-rank1} and \ref{lem-locc-epsnet}, we get that
\begin{equation}
\forall \calM \in \MM, \; I(Z;U) < n - \frac{\nya}{2} + \frac{n}{25} + O(1) + \frac{t_0\nya}{2^{\nya/2}} + u\cdot O(\sqrt{\log n}).
\end{equation}
This proves the claim.  $\square$


\subsection{1-pass LOCC strategies with $q$-outcome measurements}
\label{sec-triceratops-loccq}

In this section, we consider a more general class of 1-pass LOCC strategies, that use $q$-outcome measurements (for any constant $q$).  Again we show that the states $\ket{E(u)}$ ($u \in \set{0,1}^{\nya}$) cannot be perfectly distinguished by such an adversary.  Quantitatively, we show that such an adversary can extract at most $\approx (0.71)\nya$ bits of information about $U$; we do not believe this bound is optimal, but it does nonetheless show that a constant fraction of the information in $U$ is hidden from the adversary.  

We use a different proof technique from the previous section:  here we show an explicit lower bound on the Renyi collision entropy of $U$ conditioned on the output $Z_{\leq m}$ of the first $m$ steps of the adversary.  This ``collision entropy'' proof is a useful alternative to the ``entropy chaining'' proof of the previous section.  The collision entropy proof works quite well when $q$ is large, whereas the entropy chaining proof has some difficulty because the number of possible measurement strategies grows rapidly with $q$.  However, the collision entropy proof does not give a tight bound for any value of $q$, while the entropy chaining approach does give a tight bound when $q=2$.

\begin{theorem}\label{thm-flounder}
Let $\MM$ be the set of all 1-pass LOCC strategies in the isolated qubits model, using q-outcome measurements.  Then, with probability $\geq 1 - e^{-\Omega(n)}$ (over the choice of $E$), the following statement holds:
\begin{equation}
\forall \calM \in \MM, \; I(Z;U) \leq (0.7067)n + O(1) + O(\lg(qn)), 
\end{equation}
where $Z$ denotes the output of the strategy $\calM$.
\end{theorem}

\noindent
Proof:  First, let $L$ be an $\eps$-net for the set of single-qubit measurements with $q$ outcomes where all POVM elements have rank 1, as described in Lemma \ref{lem-epsnet-q}; and set $\eps = 1/(200qn)$.  Let $\MM'$ be the set of all strategies that use measurements chosen from the set $L$.  By Lemmas \ref{lem-locc-rank1} and \ref{lem-locc-epsnet}, any strategy in $\MM$ can be approximated by one in $\MM'$.

We will analyze the first $m$ steps of any strategy in $\MM'$, where $m = \lfloor \nya/\lg(8/3) \rfloor \approx (0.7067)\nya$.  We will show the following bound:  

\begin{lemma}\label{lem-flatfish}
With probability $\geq 1 - e^{-\Omega(n)}$ (over the choice of $E$), 
\begin{equation}\label{eq-flatfish}
\forall \calM \in \MM', \; \forall z_{1,\ldots,m}\in[q]^m, \;
H_2(U | Z_{1,\ldots,m}=z_{1,\ldots,m}) \geq \nya - m\lg(\tfrac{3}{2}) - \lg(O(qn\lg(qn))),
\end{equation}
where $Z_{1,\ldots,m}$ denotes the output of the first $m$ steps of the strategy $\calM$, $H_2$ denotes the Renyi collision entropy, and $\lg(\tfrac{3}{2}) \approx 0.5850$.
\end{lemma}

\noindent
Proof (of Lemma \ref{lem-flatfish}):  Note that the lemma is equivalent to the following statement:  with high probability (over the choice of $E$), 
\begin{verse}
for all subsets of qubits $A \subset [n]$, of size $\abs{A} = m$, \\
for all possible measurement outcomes $M_A$ that correspond to measuring the qubits in the set $A$ using any measurements in the set $L$, \\
$H_2(U | M_A)$ (where we condition on observing the measurement outcome $M_A$) is large.
\end{verse}
Note that a measurement outcome $M_A$ is uniquely represented by a rank-1 POVM element of the form $M_A = \Tensor_{i\in A} M_i$, where each $M_i$ is a POVM element acting on qubit $i$, that corresponds to one possible outcome of some measurement in the set $L$.  

We will now proceed as follows.  First, we will show that, for every $A$ and $M_A$, $\Pr[M_A]$ is approximately $\Tr(M_A)/2^m$.  Secondly, we will show that, for every $A$ and $M_A$, $\sum_u \Pr[M_A|U=u]^2$ is small.  (To show these claims, we will use large-deviation bounds for every fixed choice of $A$ and $M_A$, followed by the union bound over all $A$ and $M_A$.)  Finally, we will combine these two claims to get a lower-bound on $H_2(U|M_A)$.  

First, fix some subset of qubits $A\subset [n]$, of size $\abs{A} = m$.  Let $\rho$ be the mixed state presented to the adversary, $\rho := 2^{-\nya} \sum_{u\in\set{0,1}^{\nya}} \ket{E(u)}\bra{E(u)}$, and let $\rho_A$ be the reduced state on the subset $A$, 
\begin{equation}
\rho_A := \Tr_{[n]\setminus A}(\rho) = 2^{-\nya} \sum_{u\in\set{0,1}^{\nya}} \ket{E(u)_A}\bra{E(u)_A}, 
\text{ where } \ket{E(u)_A} := \Tensor_{a\in A} \ket{\alpha_{E(u)_a}}.  
\end{equation}
Note that $\EE_E \rho_A = I/2^m$.  We claim that, with high probability (over the choice of $E$), $\rho_A$ is close to the maximally mixed state $I/2^m$, and hence any measurement outcome $M_A$ on the subset $A$ will be observed with probability approximately $\Tr(M_A)/2^m$.  To show this, we will use the matrix Bernstein inequality \cite{tropp}.  

For convenience, define $H := \sum_{u\in\set{0,1}^{\nya}} H_u$, where $H_u := \ket{E(u)}_A\bra{E(u)}_A$.  Note that $\EE_E H = 2^{\nya-m}I$ and $\EE_E H_u = 2^{-m}I$.  Note that the $H_u$ are bounded:
\begin{equation}
\norm{H_u - \EE_E H_u} = \max\set{1-2^{-m}, 2^{-m}} < 1 =: R.
\end{equation}
The variance of $H$ is described by
\begin{equation}
\begin{split}
\EE_E[(H - \EE_E H)^2] &= \EE_E[H^2] - (\EE_E H)^2 \\
 &= \EE_E[\sum_u H_u^2 + \sum_{u\neq v} H_u H_v] - 4^{\nya-m}I \\
 &= \sum_u \EE_E[H_u^2] + \sum_{u\neq v} \EE_E[H_u] \EE_E[H_v] - 4^{\nya-m}I \\
 &= 2^{\nya-m}I + (4^{\nya}-2^{\nya}) 4^{-m}I - 4^{\nya-m}I \\
 &= 2^{\nya-m}(1-2^{-m})I.
\end{split}
\end{equation}
In particular, note that 
\begin{equation}
\sigma^2 := \norm{\EE_E[(H - \EE_E H)^2]} < 2^{\nya-m}.
\end{equation}
Then the matrix Bernstein inequality \cite{tropp} implies that, for any $t>0$, 
\begin{equation}
\Pr_E[\norm{H - \EE_E H} \geq t] \leq 2\cdot 2^m \exp\biggl( -\frac{t^2}{2(\sigma^2+\tfrac{1}{3}Rt)} \biggr)
= 2\cdot 2^m \exp\biggl( -\frac{t^2}{2(2^{\nya-m}+\tfrac{1}{3}t)} \biggr).
\end{equation}
Now set 
\begin{equation}
t := t_0\sqrt{m}2^{(\nya-m)/2}, \text{ for any } t_0 \geq 1.
\end{equation}
Recall that $0 \leq m \leq \nya - \lg\nya$.  This implies that $\lg\nya \leq \nya-m$, hence $t < t_0 \sqrt{\nya} 2^{(\nya-m)/2} \leq t_0 2^{\nya-m}$, and hence $2^{\nya-m}+\tfrac{1}{3}t \leq \tfrac{4}{3} t_0 2^{\nya-m}$.  Substituting into the above equation, we get that
\begin{equation}
\Pr_E[\norm{H - \EE_E H} \geq t_0\sqrt{m}2^{(\nya-m)/2}] \leq 2\cdot 2^m \exp(-\tfrac{3}{8} t_0 m).
\end{equation}
Recall that $\rho_A = 2^{-\nya} H$, hence this implies a large-deviation bound for $\rho_A$:
\begin{equation}
\Pr_E[\norm{\rho_A - 2^{-m}I} \geq 2^{-\nya} t_0\sqrt{m}2^{(\nya-m)/2}] \leq 2\cdot 2^m \exp(-\tfrac{3}{8} t_0 m).
\end{equation}

Now use the union bound over all subsets $A\subset[n]$ of size $\abs{A}=m$.  (There are $\binom{n}{m} < 2^n$ such sets.)  So with probability $\geq 1 - 2^{n+m+1} \exp(-\tfrac{3}{8} t_0 m)$ (over the choice of $E$), we have that 
\begin{equation}\label{eq-treefrog}
\text{for all subsets $A$ of size $m$, } \norm{\rho_A - 2^{-m}I} \leq 2^{-\nya} t_0\sqrt{m}2^{(\nya-m)/2} = 2^{-m} 2^{-(\nya-m)/2} t_0 \sqrt{m}.
\end{equation}
By setting $t_0$ to be a sufficiently large constant, we can make the failure probability exponentially small in $n$.  Finally, equation (\ref{eq-treefrog}) implies that, for any subset $A$ of size $m$, and any measurement outcome $M_A$, the probability of observing $M_A$ (which is given by $\Pr[M_A] = \Tr(M_A\rho_A)$) satisfies the bound 
\begin{equation}\label{eq-toad1}
\abs{\Pr[M_A] - 2^{-m}\Tr(M_A)} \leq 2^{-m}\Tr(M_A) 2^{-(\nya-m)/2} t_0 \sqrt{m}.
\end{equation}

Next, fix some subset of qubits $A\subset[n]$, $\abs{A}=m$, and some measurement outcome $M_A$.  We will use Bernstein's inequality to upper-bound the quantity 
\begin{equation}
F := \sum_{u\in\set{0,1}^{\nya}} F_u, \quad 
\text{where } F_u := (\Tr(M_A))^{-2} \Pr[M_A|U=u]^2.  
\end{equation}
Recall that $M_A$ is a tensor product of rank-1 operators acting on single qubits, and so $M_A/\Tr(M_A)$ can be written in the form
\begin{equation}
\tfrac{M_A}{\Tr(M_A)} = \ket{\psi_A}\bra{\psi_A}, \quad 
\text{where } \ket{\psi_A} = \Tensor_{a\in A} \ket{\psi_a}, \; \ket{\psi_a} \in \CC^2, \; 
\braket{\psi_a}{\psi_a} = 1.
\end{equation}
So we can write 
\begin{equation}
F_u = \Bigl( \bra{E(u)_A} \, \tfrac{M_A}{\Tr(M_A)} \, \ket{E(u)_A} \Bigr)^2
= \abs{\braket{\psi_A}{E(u)_A}}^4.
\end{equation}

First, we will calculate $\EE_E F = \sum_u \EE_E F_u$.  Note that $\EE_E F_u = \prod_{a\in A} \abs{\braket{\psi_a}{E(u)_a}}^4$.  We can upper-bound this as follows:
\begin{equation}\label{eq-hamster-a}
\EE_E\bigl[ \abs{\braket{\psi_a}{E(u)_a}}^4 \bigr] 
= \tfrac{1}{4}\bigl[ \abs{\braket{\psi_a}{\alpha_{00}}}^4 + \abs{\braket{\psi_a}{\alpha_{01}}}^4 + \abs{\braket{\psi_a}{\alpha_{10}}}^4 + \abs{\braket{\psi_a}{\alpha_{11}}}^4 \bigr]
= \tfrac{1}{4} \bra{\psi_a}^{\tensor 2} T \ket{\psi_a}^{\tensor 2}, 
\end{equation}
where we define the matrix $T \in \CC^{4\times 4}$ to be 
\begin{equation}
T := 
\Bigl(\ket{0}\bra{0}\Bigr)^{\tensor 2} + \Bigl(\ket{1}\bra{1}\Bigr)^{\tensor 2} + 
\Bigl(\ket{+}\bra{+}\Bigr)^{\tensor 2} + \Bigl(\ket{-}\bra{-}\Bigr)^{\tensor 2}. 
\end{equation}
Now write the spectral decomposition of $T$:
\begin{equation}
T = \ket{\Psi^+}\bra{\Psi^+} + 2\ket{\Phi^+}\bra{\Phi^+} + \ket{\Phi^-}\bra{\Phi^-},
\end{equation}
where $\ket{\Psi^\pm} = \tfrac{1}{\sqrt{2}}(\ket{01}\pm\ket{10})$ and $\ket{\Phi^\pm} = \tfrac{1}{\sqrt{2}}(\ket{00}\pm\ket{11})$ are the Bell states.  Now write $\ket{\psi_a}$ in the form $\ket{\psi_a} = \alpha\ket{0} + \beta\ket{1}$.  This implies
\begin{equation}
\ket{\psi_a}^{\tensor 2} = \sqrt{2}\alpha\beta\ket{\Psi^+} + \tfrac{1}{\sqrt{2}}(\alpha^2+\beta^2)\ket{\Phi^+} + \tfrac{1}{\sqrt{2}}(\alpha^2-\beta^2)\ket{\Phi^-}.
\end{equation}
Now we calculate
\begin{equation}\label{eq-hamster-z}
\begin{split}
\bra{\psi_a}^{\tensor 2} (2T) \ket{\psi_a}^{\tensor 2}
&= 4\abs{\alpha}^2\abs{\beta}^2 + 2(\alpha^2+\beta^2)^*(\alpha^2+\beta^2) + (\alpha^2-\beta^2)^*(\alpha^2-\beta^2) \\
&= 4\abs{\alpha}^2\abs{\beta}^2 + 3\abs{\alpha}^4 + (\alpha^2)^*\beta^2 + (\beta^2)^*\alpha^2 + 3\abs{\beta}^4 \\
&= 2(\abs{\alpha}^2+\abs{\beta}^2)^2 + \abs{\alpha^2+\beta^2}^2 \\
&\leq 3(\abs{\alpha}^2+\abs{\beta}^2)^2 = 3.
\end{split}
\end{equation}
This implies that $\EE_E\bigl[ \abs{\braket{\psi_a}{E(u)_a}}^4 \bigr] \leq \tfrac{3}{8}$, and hence $\EE_E F_u \leq (\tfrac{3}{8})^m$ and $\EE_E F \leq 2^{\nya}(\tfrac{3}{8})^m$.  

In addition, we bound the variance of $F_u$ as follows (using the fact that $0\leq F_u\leq 1$):  $\Var_E F_u \leq \EE_E (F_u^2) \leq \EE_E F_u \leq (\tfrac{3}{8})^m$.  

Now Bernstein's inequality \cite{dubhashi} implies that, for all $t>0$, 
\begin{equation}
\Pr_E[F > \EE_E F + t] \leq \exp\biggl( -\frac{t^2}{2\cdot 2^{\nya} (\tfrac{3}{8})^m + \tfrac{2}{3} t} \biggr).
\end{equation}
Recall that $m\leq \nya/\lg(\tfrac{8}{3})$, and note that this implies $2^{\nya} (\tfrac{3}{8})^m \geq 1$.  Now set 
\begin{equation}
t = t_1 [2^{\nya} (\tfrac{3}{8})^m]^{1/2}, \text{ for any } t_1 \geq 1.  
\end{equation}
This implies $2\cdot 2^{\nya} (\tfrac{3}{8})^m + \tfrac{2}{3} t \leq (2+\tfrac{2}{3}) t_1 2^{\nya} (\tfrac{3}{8})^m$.  Substituting into the above equation, we get 
\begin{equation}
\Pr_E[F > 2^{\nya} (\tfrac{3}{8})^m + t_1 [2^{\nya} (\tfrac{3}{8})^m]^{1/2}] \leq \exp(-\tfrac{3}{8}t_1).
\end{equation}

Now take the union bound over all subsets $A\subset[n]$ of size $\abs{A}=m$, and all measurement outcomes $M_A$ that correspond to measurements chosen from the set $L$ and performed on the qubits in the set $A$.  (There are $\binom{n}{m} < 2^n$ such sets, and $(q\abs{L})^m \leq (q O(1/\eps)^{3q})^m \leq (q O(qn)^{3q})^m \leq 2^{O(qn\lg(qn))}$ such measurement outcomes.)  Then, with probability $\geq 1 - 2^{O(qn\lg(qn))} \exp(-\tfrac{3}{8}t_1)$, we have that 
\begin{equation}\label{eq-toad2}
\text{for all subsets $A$ of size $m$, and all measurement outcomes $M_A$, } 
F \leq 2^{\nya} (\tfrac{3}{8})^m + t_1 [2^{\nya} (\tfrac{3}{8})^m]^{1/2}.
\end{equation}
By setting $t_1 := \Theta(qn\lg(qn))$, we can make the failure probability exponentially small in $n$.  

Finally, we will combine equations (\ref{eq-toad1}) and (\ref{eq-toad2}) to get a lower bound on $H_2(U|M_A)$.  For any $A$ and $M_A$, we write 
\begin{equation}
\begin{split}
2^{-H_2(U|M_A)} &= \sum_{u\in\set{0,1}^{\nya}} \Pr[U=u|M_A]^2 \\
 &= \Pr[M_A]^{-2} 4^{-\nya} \sum_{u\in\set{0,1}^{\nya}} \Pr[M_A|U=u]^2 \\
 &= \Pr[M_A]^{-2} 4^{-\nya} (\Tr(M_A))^2 F \\
 &\leq [2^{-m} \Tr(M_A) (1 - 2^{-(\nya-m)/2} t_0 \sqrt{m})]^{-2} 4^{-\nya} (\Tr(M_A))^2 [2^{\nya} (\tfrac{3}{8})^m + t_1 [2^{\nya} (\tfrac{3}{8})^m]^{1/2} ] \\
 &\leq 4^m (1 - 2^{-(\nya-m)/2} t_0 \sqrt{m})^{-2} 4^{-\nya} 2^{\nya} (\tfrac{3}{8})^m (1+t_1) \\
 &\leq 2^{-\nya} (\tfrac{3}{2})^m O(qn\lg(qn)).
\end{split}
\end{equation}
This implies
\begin{equation}
\begin{split}
H_2(U|M_A) &\geq \nya - m\lg(\tfrac{3}{2}) - \lg(O(qn\lg(qn))) \\
 &\approx \nya - (0.5850)m - \lg(O(qn\lg(qn))).
\end{split}
\end{equation}
This completes the proof of Lemma \ref{lem-flatfish}.  $\square$

{\vskip 11pt}

We now return to the proof of Theorem \ref{thm-flounder}.  Consider any strategy $\calM \in \MM'$.  We want to bound the mutual information 
\begin{equation}
I(U;Z) = I(U; Z_{1,\ldots,m}) + I(U; Z_{m+1,\ldots,n} | Z_{1,\ldots,m}).
\end{equation}

We bound the first term using Lemma \ref{lem-flatfish}.  First we write 
\begin{equation}
\begin{split}
I(U; Z_{1,\ldots,m}) &= H(U) - H(U | Z_{1,\ldots,m}) \\
 &\leq \nya - \sum_{z_{1,\ldots,m}\in[q]^m} \Pr[Z_{1,\ldots,m}=z_{1,\ldots,m}] H(U | Z_{1,\ldots,m}=z_{1,\ldots,m}).
\end{split}
\end{equation}
For any particular string of measurement outcomes $z_{1,\ldots,m} \in [q]^m$, let $A \subset [n]$ be the set of qubits that were measured, and let $M_A$ be the corresponding POVM element.  Then, by Lemma \ref{lem-flatfish}, we have 
\begin{equation}
\begin{split}
H(U | Z_{1,\ldots,m}=z_{1,\ldots,m}) &\geq H_2(U | Z_{1,\ldots,m}=z_{1,\ldots,m}) \\
 &\geq \nya - \nya \tfrac{\lg(3/2)}{\lg(8/3)} - O(\lg(qn)).
\end{split}
\end{equation}

We bound the second term using Holevo's inequality \cite{NC}.  First we write 
\begin{equation}
I(U; Z_{m+1,\ldots,n} | Z_{1,\ldots,m})
 = \sum_{z_{1,\ldots,m}\in[q]^m} \Pr[Z_{1,\ldots,m}=z_{1,\ldots,m}] I(U; Z_{m+1,\ldots,n} | Z_{1,\ldots,m}=z_{1,\ldots,m}).
\end{equation}
For any particular string of measurement outcomes $z_{1,\ldots,m} \in [q]^m$, let $A \subset [n]$ be the set of qubits that were measured, and let $p(u) = \Pr[U=u | Z_{1,\ldots,m}=z_{1,\ldots,m}]$.  Then we have 
\begin{equation}
\begin{split}
I(U; Z_{m+1,\ldots,n} | Z_{1,\ldots,m}=z_{1,\ldots,m}) 
 &\leq S\biggl( \sum_{u \in \set{0,1}^{\nya}} p(u) \ket{E(u)}\bra{E(u)}_{[n] \setminus A} \biggr) 
     - \sum_{u \in \set{0,1}^{\nya}} p(u) S\Bigl( \ket{E(u)}\bra{E(u)}_{[n] \setminus A} \Bigr) \\
 &\leq n-m \leq n - \tfrac{\nya}{\lg(8/3)} + 1.
\end{split}
\end{equation}

Combining these bounds, we get 
\begin{equation}
\begin{split}
I(U;Z) &\leq \nya - \nya + \nya \tfrac{\lg(3/2)}{\lg(8/3)} + O(\lg(qn)) + n - \tfrac{\nya}{\lg(8/3)} + 1 \\
 &\leq n - \nya \tfrac{\lg(4/3)}{\lg(8/3)} + O(\lg(qn)) \\
 &\approx n - (0.2933)\nya + O(\lg(qn)).
\end{split}
\end{equation}
$\square$


\section{One-time memories from isolated qubits}
\label{sec-stegosaurus}

A one-time memory (OTM) is a device that implements the following functionality \cite{GKR}:  one party (Alice) can write two messages $s,t\in\set{0,1}^k$ into the device, and then give the device to another party (Bob); after receiving the device, Bob can then choose to read either $s$ or $t$, but not both.  The main application of OTM's is to construct one-time programs \cite{GKR,goyal,bellare}.

In this section, we will construct devices which are similar to OTM's, in the isolated qubits model.  Our devices will not implement the ideal OTM functionality described above, but we conjecture that they do provide a weaker ``leaky OTM'' functionality that is still sufficient to construct one-time programs.  We will first define this ``leaky OTM'' functionality, and then describe our construction based on isolated qubits.  

\subsection{Leaky OTM's}
\label{sec-stegosaurus-leaky}

\begin{definition}
Fix some class of adversary strategies $\MM$, some leakage parameter $\delta \in [0,1]$, and some failure probability $\eps \in [0,1]$.  A \emph{leaky one-time memory (leaky OTM)} with parameters $(\MM,\delta,\eps)$ is a device that has the following behavior.  Suppose that the device is programmed with two messages $s$ and $t$ chosen uniformly at random in $\set{0,1}^k$; and let $S$ and $T$ be the random variables containing these messages.  Then:
\begin{enumerate}
\item Correctness:  There exists an honest strategy $\calM^{(1)} \in \MM$ that interacts with the device and recovers the message $s$ with probability $\geq 1-\eps$.  Likewise, there exists an honest strategy $\calM^{(2)} \in \MM$ that recovers the message $t$ with probability $\geq 1-\eps$.  
\item Leaky security:  For every strategy $\calM \in \MM$, if $Z$ is the random variable containing the classical information output by $\calM$, then with probability $\geq 1-\eps$, $Z$ takes on a value $z$ such that $H_\infty^\eps(S,T|Z=z) \geq (1-\delta)k$.  (Here $H_\infty^\eps$ is the smoothed min-entropy.)
\end{enumerate}
\end{definition}

This leaky OTM is weaker than the ideal OTM in two important respects:  it assumes that the messages $s$ and $t$ are chosen uniformly at random, independent of all other variables; and it allows the adversary to obtain partial information about both $s$ and $t$, so long as the adversary still has $(1-\delta)k$ bits of uncertainty (as measured by the smoothed min-entropy).

Also, note that when we choose $\MM$ to be the class of LOCC adversaries, the above definition yields information-theoretic (rather than computational) notions of correctness and security.  In particular, the honest strategies are not required to be computationally efficient, but at the same time, security holds against LOCC adversaries with unbounded computational power.

We remark that this definition is mostly classical, rather than quantum.  In particular, this definition assumes that the party who programs the OTM is classical, so that the messages $s$ and $t$ are classical bit strings.  This definition also assumes that the party who reads the OTM may make (quantum) LOCC measurements, but immediately outputs a classical string $z$.  These assumptions are reasonable, since the isolated qubits model is mostly classical, as LOCC operations can never generate quantum entanglement.  

We conjecture that one can construct one-time programs from (computationally-efficient) leaky OTM's, provided the leakage parameter $\delta$ is a sufficiently small constant, and the failure probability $\eps$ is negligible.  To see why this is plausible, consider the one-time programs in \cite{GKR}, which are based on Yao's garbled circuits.  Here, the OTM's contain keys for an encryption scheme.  These keys are chosen independently at random, and are never re-used.  Furthermore, one can use a leak-resistant encryption scheme, which can tolerate leakage of a constant fraction of the bits of the key \cite{AGV}.  Here, the adversary's remaining uncertainty about the key is expressed using the (smoothed) min-entropy.  This suggests that leaky OTM's will be sufficient for this construction.

\subsection{Construction using isolated qubits}
\label{sec-stegosaurus-construction}

We consider a system of $n$ isolated qubits, and we construct a set of states $\ket{E(s,t)}$ (for $s,t\in\set{0,1}^k$), where $k \approx (0.3991)n$.  First, choose a random function $C: \set{0,1}^k \rightarrow \set{0,1}^n$, i.e., for each $s \in \set{0,1}^k$, choose $C(s) \in \set{0,1}^n$ independently and uniformly at random.  Similarly, choose a random function $D: \set{0,1}^k \rightarrow \set{0,1}^n$.  Now define 
\begin{equation}
\ket{E(s,t)} := \Tensor_{a=1}^n \ket{\alpha_{C(s)_a D(t)_a}}, 
\end{equation}
where the single-qubit states $\ket{\alpha_{00}}$, $\ket{\alpha_{01}}$, $\ket{\alpha_{10}}$, $\ket{\alpha_{11}}$ are defined in the same way as in the previous section:
\begin{equation}
\ket{\alpha_{00}} = \ket{0}, \quad
\ket{\alpha_{11}} = \ket{1}, \quad
\ket{\alpha_{01}} = \tfrac{1}{\sqrt{2}} (\ket{0}+\ket{1}), \quad
\ket{\alpha_{10}} = \tfrac{1}{\sqrt{2}} (\ket{0}-\ket{1}).
\end{equation}
Note that these states can be prepared using single-qubit operations in the isolated qubits model.

We now describe the ``honest'' measurement strategies, that can be used to learn either $s$ or $t$.  The states $\ket{E(s,t)}$ are reminiscent of Wiesner's conjugate coding \cite{wiesner}, in that measuring in one basis reveals information about $s$, while measuring in another basis reveals information about $t$.  Let us define the states 
\begin{equation}
\ket{\beta_\phi} := \cos(\phi)\ket{0} + \sin(\phi)\ket{1}, \quad \phi \in \RR.
\end{equation}
Then measuring each qubit in the basis $\set{\ket{\beta_{\pi/8}}, \ket{\beta_{5\pi/8}}}$ returns a ``noisy'' copy of the string $C(s)$, which can be decoded to recover $s$ (since, with high probability, $C$ is a good error-correcting code).  Likewise, measuring each qubit in the basis $\set{\ket{\beta_{-\pi/8}}, \ket{\beta_{3\pi/8}}}$ returns a ``noisy'' copy of the string $D(t)$, which can be decoded to recover $t$.

Is there some measurement strategy that will reveal both $s$ and $t$?  Wiesner pointed out that there does exist a joint measurement on the $n$ qubits that can recover both $s$ and $t$; this also follows from the ``gentle measurement lemma'' of Winter \cite{gentle}.  However, when the codes $C$ and $D$ are ``unstructured,'' one may expect this measurement to be highly entangled, hence impossible to perform using only LOCC operations.  We will give rigorous evidence that this is indeed the case.

We will show that these states partially satisfy the definition of a leaky OTM.  In particular, with high probability over the random choice of $C$ and $D$, the following statements hold:
\begin{enumerate}
\item Suppose $s$ and $t$ are chosen uniformly at random.  There exists a sequence of single-qubit projective measurements that can reconstruct $s$ with probability $\geq 1 - e^{-\Omega(n)}$.  Likewise, there exists a sequence of single-qubit projective measurements that can reconstruct $t$ with probability $\geq 1 - e^{-\Omega(n)}$.
\item Suppose $s$ and $t$ are chosen uniformly at random.  No 1-pass LOCC measurement strategy using 2-outcome measurements can recover more than $\approx (1.9190)k$ bits of information about $s$ and $t$.  That is, if $S$ and $T$ are the random variables containing the strings $s$ and $t$, and $Z$ is the output of the measurement strategy, then 
\begin{equation}
I(Z;S,T) \leq (1.9190)k + O(\sqrt{n} \log n).  
\end{equation}
Note that we can write $I(Z;S,T) = H(S,T) - H(S,T|Z)$, hence this implies 
\begin{equation}
H(S,T|Z) \geq (0.081)k - O(\sqrt{n} \log n).  
\end{equation}
\end{enumerate}

These statements are similar to the definition of a leaky OTM, where $\MM$ is the set of 1-pass LOCC measurement strategies using 2-outcome measurements, the leakage parameter is $\delta \approx 0.9190$, the failure probability is $\eps = e^{-\Omega(n)}$, and the security condition is relaxed by replacing the smoothed min-entropy $H_\infty^\eps$ with the Shannon entropy $H$.  (Also, we set aside the requirement that the honest strategies must be computationally efficient.)

We believe it should be possible to prove much stronger results of this type.  In particular, it should be possible to improve our bound on the leakage parameter $\delta$, as our current proof technique is somewhat ad hoc.  Also, we remark that some parts of our proof already imply bounds on the smoothed min-entropy $H_\infty^\eps$.  Specifically, in Lemma \ref{lem-cuttlefish}, we actually lower-bound the collision entropy $H_2(S,T|Z_{1,\ldots,m}=z_{1,\ldots,m})$, where $Z_{1,\ldots,m}$ represents the measurement outcomes of the first $m$ qubits measured by the adversary, where $m \approx (0.7067)k$.  This directly implies a lower-bound on the smoothed min-entropy $H_\infty^\eps(S,T|Z_{1,\ldots,m}=z_{1,\ldots,m})$, using a standard argument.\footnote{In particular, for any $\theta \geq 0$, if we set $\eps = 2^{-\theta}$, then $H_\infty^\eps(X) \geq H_2(X) - \theta$; this follows from Markov's inequality.}  


\subsection{Correctness for honest parties}
\label{sec-stegosaurus-correct}

First, we show that the honest strategies for recovering either $s$ or $t$ (as described above) do succeed with high probability.  Without loss of generality, suppose we want to recover $s$.  Let $S$ and $T$ be random variables, distributed independently and uniformly on $\set{0,1}^k$.  We are given the state $\ket{E(S,T)}$, and we measure each qubit in the basis $\set{\ket{\beta_{\pi/8}}, \ket{\beta_{5\pi/8}}}$.  Let $Z$ be the random variable containing the string of measurement outcomes, i.e., $Z$ takes values in $\set{0,1}^n$.  

We decode $Z$ as follows:  we output any string $t \in \set{0,1}^k$ such that $d_H(C(t),Z) \leq r$, where $d_H$ denotes the Hamming distance, and $r$ is a parameter that we will set below.  If there are multiple candidate strings $t$, we pick one of them in some arbitrary fashion.  Let $\Shat$ be the random variable containing the output of this procedure.  

Observe that $Z$ is the output of a binary symmetric channel $BSC(p_e)$ applied to the string $C(S)$, where the error probability $p_e$ is given by 
\begin{equation}
p_e := \sin^2(\pi/8) \approx 0.1464.
\end{equation}
Recall that the channel $BSC(p_e)$ has capacity $1-h(p_e) \approx 0.3991$, where $h(p)$ is the binary entropy function, 
\begin{equation}
h(p) := -p\lg(p) - (1-p)\lg(1-p).
\end{equation}
Also, note that the expected number of errors introduced by the channel is $np_e$.  

This suggests that we should set the parameters $k$ and $r$ as follows:
\begin{equation}\label{eq-penguin}
k := n(1-h(p_e)-\theta), 
\end{equation}
\begin{equation}
r := n(p_e+\tau), 
\end{equation}
where $0<\theta \ll 1$ and $0<\tau \ll 1$ are small constants.

We prove the following statement, which is essentially Shannon's noisy coding theorem for the binary symmetric channel, using an argument from \cite{welsh}.  This shows that, when we choose $\theta$ and $\tau$ appropriately, then with high probability over the choice of the random code $C$, $\Pr[\Shat=S]$ is close to 1.

\begin{proposition}\label{prop-sea-cucumber}
Fix any constants $\lambda \geq 1$, $0 < \tau \leq \tfrac{1}{2} - p_e \approx 0.3536$, and $\theta > \tau h'(p_e) \approx (2.5431) \tau$.  Then for all sufficiently large $n$, the following statement holds:  with probability $\geq 1-\tfrac{1}{\lambda}$ (over the choice of $C$), we have 
\begin{equation}
\Pr[\Shat=S] \geq 1 - \lambda \bigl[ e^{-2\tau^2 n} + 2^{-n (\theta - \tau h'(p_e))} \bigr]
\geq 1 - \lambda e^{-\Omega(n)}.
\end{equation}
\end{proposition}

\noindent
Proof:  See Appendix \ref{app-noisy-coding}.

\subsection{Security against 1-pass LOCC adversaries using 2-outcome measurements}
\label{sec-stegosaurus-secure}

In this section we will upper-bound the amount of information that can be extracted from our OTM devices by any 1-pass LOCC adversary using 2-outcome measurements.  We use the collision entropy technique to analyze the first few steps taken by the adversary; then we use the entropy chaining technique to prove bounds on the adversary's subsequent steps.  To deal with the correlations among the states $\ket{E(s,t)}$, we will use large deviation bounds for sums of locally dependent random variables \cite{janson, dubhashi}.

\begin{theorem}\label{thm-squid}
Let $\MM$ be the set of all 1-pass LOCC strategies that use 2-outcome measurements.  Then, with probability $\geq 1 - e^{-\Omega(n)}$ (over the choice of $C$ and $D$), the following statement holds:
\begin{equation}
\forall \calM \in \MM, \; I(Z;S,T) \leq (1.9190)k + O(\sqrt{n} \log n), 
\end{equation}
where $Z$ denotes the output of the strategy $\calM$.  Equivalently, we can write:
\begin{equation}
\forall \calM \in \MM, \; H(S,T|Z) \geq (0.081)k - O(\sqrt{n} \log n).
\end{equation}
\end{theorem}

\noindent
Proof:  First, let $L$ be an $\eps$-net for the set of single-qubit measurements with 2 outcomes where all POVM elements have rank 1, as described in Lemma \ref{lem-epsnet-2}; and set $\eps = 1/(200n)$.  Let $\MM'$ be the set of all strategies that use measurements chosen from the set $L$.  By Lemmas \ref{lem-locc-rank1} and \ref{lem-locc-epsnet}, any strategy in $\MM$ can be approximated by one in $\MM'$.

We will analyze the first $m$ steps of any strategy in $\MM'$, where 
\begin{equation}\label{eq-shrimp0}
m := \lfloor k/\lg(8/3) \rfloor \approx (0.7067)k.  
\end{equation}
We will show the following bound:  

\begin{lemma}\label{lem-cuttlefish}
With probability $\geq 1 - e^{-\Omega(n)}$ (over the choice of $C$ and $D$), 
\begin{equation}\label{eq-cuttlefish}
\forall \calM \in \MM', \; \forall z_{1,\ldots,m}\in\set{0,1}^m, \;
H_2(S,T | Z_{1,\ldots,m}=z_{1,\ldots,m}) \geq 2k - m\lg(\tfrac{3}{2}) - \lg(O(n\lg n)),
\end{equation}
where $Z_{1,\ldots,m}$ denotes the output of the first $m$ steps of the strategy $\calM$, $H_2$ denotes the Renyi collision entropy, and $\lg(\tfrac{3}{2}) \approx 0.5850$.
\end{lemma}

\noindent
Remark:  Equation (\ref{eq-cuttlefish}) is equivalent to the following statement:  
\begin{verse}
for all subsets of qubits $A \subset [n]$, of size $\abs{A} = m$, \\
for all possible measurement outcomes $M_A$, that can be obtained by measuring the qubits in $A$ (using measurements chosen from the set $L$), \\
$H_2(S,T | M_A) \geq 2k - m\lg(\tfrac{3}{2}) - \lg(O(n\lg n))$. 
\end{verse}
Recall that a measurement outcome $M_A$ is uniquely represented by a rank-1 POVM element of the form $M_A = \Tensor_{i\in A} M_i$, where each $M_i$ is a POVM element acting on qubit $i$, that corresponds to one possible outcome of some measurement in the set $L$.  

{\vskip 11pt}

\noindent
Proof (of Lemma \ref{lem-cuttlefish}):  First, we will show that, for every $A$ and $M_A$, $\Pr[M_A]$ is approximately $\Tr(M_A)/2^m$.  Secondly, we will show that, for every $A$ and $M_A$, $\sum_u \Pr[M_A|S=s,T=t]^2$ is small.  (To show these claims, we will use large-deviation bounds for every fixed choice of $A$ and $M_A$, followed by the union bound over all $A$ and $M_A$.)  Finally, we will combine these two claims to get a lower-bound on $H_2(S,T|M_A)$.  

First, fix some subset of qubits $A\subset [n]$, of size $\abs{A} = m$, and fix some measurement outcome $M_A$.  Let $\rho$ be the mixed state presented to the adversary, $\rho := 4^{-k} \sum_{s,t\in\set{0,1}^k} \ket{E(s,t)}\bra{E(s,t)}$, and let $\rho_A$ be the reduced state on the subset $A$, 
\begin{equation}
\rho_A := \Tr_{[n]\setminus A}(\rho) = 4^{-k} \sum_{s,t\in\set{0,1}^k} \ket{E(s,t)_A}\bra{E(s,t)_A}, 
\text{ where } \ket{E(s,t)_A} := \Tensor_{a\in A} \ket{\alpha_{C(s)_a D(t)_a}}.  
\end{equation}
Recall that $M_A$ is a tensor product of rank-1 operators acting on single qubits.  Moreover, since the adversary uses single-qubit measurements with 2 outcomes, each measurement outcome is a rank-1 projector.  So $\Tr(M_A)=1$, and $M_A$ can be written in the form
\begin{equation}
M_A = \ket{\psi_A}\bra{\psi_A}, \quad 
\text{where } \ket{\psi_A} = \Tensor_{a\in A} \ket{\psi_a}, \; \ket{\psi_a} \in \CC^2, \; 
\braket{\psi_a}{\psi_a} = 1.
\end{equation}

We will use Bernstein's inequality for locally dependent random variables \cite{janson} to lower-bound the quantity 
\begin{equation}
\Pr[M_A] = \Tr(M_A\rho_A) = 4^{-k} \sum_{s,t\in\set{0,1}^k} \abs{\braket{\psi_A}{E(s,t)_A}}^2.
\end{equation}
For convenience, let us define the random variables 
\begin{equation}
H := \sum_{s,t\in\set{0,1}^k} H_{st}, \quad H_{st} := \abs{\braket{\psi_A}{E(s,t)_A}}^2.
\end{equation}
We can calculate their expectation values:
\begin{equation}
\EE_{CD} H_{st} = \prod_{a\in A} \EE_{CD}[ \abs{\braket{\psi_a}{E(s,t)_a}}^2 ] = 2^{-m},
\end{equation}
hence $\EE_{CD} H = 4^k 2^{-m}$.  We can also bound their variances:
\begin{equation}
\Var_{CD} H_{st} \leq \EE_{CD}[H_{st}^2] = \prod_{a\in A} \EE_{CD}[ \abs{\braket{\psi_a}{E(s,t)_a}}^4 ]
\leq (\tfrac{3}{8})^m, 
\end{equation}
where in the last step we re-used the argument shown in equations (\ref{eq-hamster-a})-(\ref{eq-hamster-z}) in the proof of Lemma \ref{lem-flatfish}.

We claim that the dependency graph $\Gamma$ of the random variables $H_{st}$ ($s,t\in\set{0,1}^k$) has chromatic number $\chi(\Gamma) \leq 2^k$.  To see this, note that two vertices $(s,t)$ and $(s',t')$ in $\Gamma$ are adjacent if and only if $s=s'$ or $t=t'$.  We can color the vertices of $\Gamma$ as follows:  assign each vertex $(s,t)$ the color specified by the string $s\oplus t \in \set{0,1}^k$ (where $\oplus$ denotes bitwise XOR).  It is easy to check that this is a legal coloring, which uses $2^k$ colors.

Using Bernstein's inequality for locally dependent random variables \cite{janson}, we get that for all $\tau>0$, 
\begin{equation}
\begin{split}
\Pr_{CD}[H \leq \EE_{CD}H - \tau] 
&\leq \exp\biggl( -\frac{8\tau^2}{25\cdot 2^k (4^k (\tfrac{3}{8})^m + \tfrac{1}{3} \tau)} \biggr) \\
&\leq \exp\biggl( -\frac{8\tau^2}{25\cdot 2^k \max\set{2\cdot (\tfrac{3}{2})^k (\tfrac{8}{3})^{k-m}, \tfrac{2}{3} \tau}} \biggr) \\
&= \max\biggl\lbrace \exp\biggl( -\frac{4\tau^2}{25\cdot 3^k (\tfrac{8}{3})^{k-m}} \biggr), \; 
\exp\biggl( -\frac{12\tau^2}{25\cdot 2^k \tau} \biggr) \biggr\rbrace.
\end{split}
\end{equation}
Now set 
\begin{equation}
\tau := 4^k 2^{-m} k^{-1} = 2^k 2^{k-m} k^{-1}, 
\end{equation}
which implies
\begin{equation}
\Pr_{CD}[H \leq 4^k 2^{-m} (1-k^{-1})] 
\leq \max\bigl\lbrace \exp(-\tfrac{4}{25} (\tfrac{4}{3})^k (\tfrac{3}{2})^{k-m} k^{-2}), \;
\exp(-\tfrac{12}{25} 2^{k-m} k^{-1}) \bigr\rbrace.
\end{equation}

Now take the union bound over all subsets $A\subset[n]$ of size $\abs{A}=m$, and all measurement outcomes $M_A$ that correspond to measurements chosen from the set $L$ and performed on the qubits in the set $A$.  (There are $\binom{n}{m} < 2^n$ such sets, and $(2\abs{L})^m \leq (O(1/\eps^2))^m \leq (O(n^2))^m \leq 2^{O(n\lg n)}$ such measurement outcomes.)  So, with probability $\geq 1 - 2^{O(n\lg n)} \cdot \max\bigl\lbrace \exp(-\tfrac{4}{25} (\tfrac{4}{3})^k (\tfrac{3}{2})^{k-m} k^{-2}), \;
\exp(-\tfrac{12}{25} 2^{k-m} k^{-1}) \bigr\rbrace$ (over the choice of $C$ and $D$), we have that: 
\begin{equation}\label{eq-goldfish1}
\begin{split}
&\text{for all subsets $A$ of size $m$, and all measurement outcomes $M_A$, } \\
&\quad\quad H \geq 4^k 2^{-m} (1-k^{-1}), \text{ and hence, } \Pr[M_A] \geq 2^{-m} (1-k^{-1}).
\end{split}
\end{equation}
Recall that $m \leq (0.7061)k$, and $k \approx (0.3991)n$; this implies that the failure probability is doubly-exponentially small in $n$.  

Next, fix some subset of qubits $A\subset[n]$, $\abs{A}=m$, and some measurement outcome $M_A = \ket{\psi_A}\bra{\psi_A}$, as before.  We will now upper-bound the quantity 
\begin{equation}
F := \sum_{s,t\in\set{0,1}^k} F_{st}, \quad 
\text{where } F_{st} := \Pr[M_A|S=s,T=t]^2 = \abs{\braket{\psi_A}{E(s,t)_A}}^4.  
\end{equation}

First, note that $\EE_{CD} F_{st} \leq (\tfrac{3}{8})^m$ and $\EE_{CD} F \leq 4^k (\tfrac{3}{8})^m$, by the same argument shown in equations (\ref{eq-hamster-a})-(\ref{eq-hamster-z}) in the proof of Lemma \ref{lem-flatfish}.  In addition, since $0\leq F_{st}\leq 1$, we have that $\Var_{CD} F_{st} \leq \EE_{CD} (F_{st}^2) \leq \EE_{CD} F_{st} \leq (\tfrac{3}{8})^m$.  

Now Bernstein's inequality for locally dependent random variables \cite{janson} implies that, for all $\tau>0$, 
\begin{equation}
\Pr_{CD}[F > \EE_{CD} F + \tau] \leq \exp\biggl( -\frac{8\tau^2}{25\cdot 2^k (4^k (\tfrac{3}{8})^m + \tfrac{1}{3} \tau)} \biggr).
\end{equation}
Recall that $m\leq k/\lg(\tfrac{8}{3})$, and note that this implies $2^k (\tfrac{3}{8})^m \geq 1$.  Now set 
\begin{equation}
\tau = \tau_1 4^k (\tfrac{3}{8})^m, \text{ for any } \tau_1 \geq 1.  
\end{equation}
This implies $4^k (\tfrac{3}{8})^m + \tfrac{1}{3} \tau \leq (1+\tfrac{1}{3}) \tau_1 4^k (\tfrac{3}{8})^m = \tfrac{4}{3} \tau$.  Substituting into the above equation, we get 
\begin{equation}
\Pr_E[F > 4^k (\tfrac{3}{8})^m (1+\tau_1)] \leq \exp(-\tfrac{6}{25} \tau_1 2^k (\tfrac{3}{8})^m)
\leq \exp(-\tfrac{6}{25} \tau_1).
\end{equation}

Now take the union bound over all subsets $A\subset[n]$ of size $\abs{A}=m$, and all measurement outcomes $M_A$ that correspond to measurements chosen from the set $L$ and performed on the qubits in the set $A$.  Then, with probability $\geq 1 - 2^{O(n\lg n)} \exp(-\tfrac{6}{25}\tau_1)$, we have that: 
\begin{equation}\label{eq-goldfish2}
\text{for all subsets $A$ of size $m$, and all measurement outcomes $M_A$, } 
F \leq 4^k (\tfrac{3}{8})^m (1+\tau_1).
\end{equation}
By setting $\tau_1 := \Theta(n\lg n)$, we can make the failure probability exponentially small in $n$.  

Finally, we will combine equations (\ref{eq-goldfish1}) and (\ref{eq-goldfish2}), to get a lower bound on $H_2(S,T|M_A)$.  For any $A$ and $M_A$, we write 
\begin{equation}
\begin{split}
2^{-H_2(S,T|M_A)} &= \sum_{s,t\in\set{0,1}^k} \Pr[S=s,T=t|M_A]^2 \\
 &= \Pr[M_A]^{-2} 4^{-2k} \sum_{s,t\in\set{0,1}^k} \Pr[M_A|S=s,T=t]^2 \\
 &= \Pr[M_A]^{-2} 4^{-2k} F \\
 &\leq [2^{-m} (1-k^{-1})]^{-2} 4^{-2k} 4^k (\tfrac{3}{8})^m (1+\tau_1) \\
 &= 4^{-k} (\tfrac{3}{2})^m \tfrac{1+\tau_1}{(1-k^{-1})^2}.
\end{split}
\end{equation}
This implies
\begin{equation}
\begin{split}
H_2(S,T|M_A) &\geq 2k - m\lg(\tfrac{3}{2}) - \lg(1+\tau_1) + 2\lg(1-k^{-1}) \\
 &\geq 2k - m\lg(\tfrac{3}{2}) - \lg(O(n\lg n)).
\end{split}
\end{equation}
This completes the proof of Lemma \ref{lem-cuttlefish}.  $\square$

{\vskip 11pt}

We now return to the proof of Theorem \ref{thm-squid}.  Consider any measurement strategy $\calM \in \MM'$, and let $Z$ be its output.  We will upper-bound the amount of information extracted during the first $m$ steps, using Lemma \ref{lem-cuttlefish}:
\begin{equation}\label{eq-shrimp1}
\begin{split}
I(S,T; Z_{1,\ldots,m}) &= H(S,T) - H(S,T|Z_{1,\ldots,m}) \\
 &\leq H(S,T) - \sum_{z_{1,\ldots,m}} \Pr[Z_{1,\ldots,m}=z_{1,\ldots,m}] H_2(S,T|Z_{1,\ldots,m}=z_{1,\ldots,m}) \\
 &\leq 2k - [2k - m\lg(\tfrac{3}{2}) - \lg(O(n\lg n))] \\
 &= m\lg(\tfrac{3}{2}) + \lg(O(n\lg n)) \\
 &= k \tfrac{\lg(3/2)}{\lg(8/3)} + \lg(O(n\lg n)).
\end{split}
\end{equation}

Next, we will analyze the subsequent steps of the adversary.  First, let us fix some subset of qubits $A\subset[n]$, of size $\abs{A}=m$, and some measurement outcome $M_A$; these represent past actions of the adversary during its first $m$ steps.  We will then upper-bound the amount of information gained by the adversary in the next $\mya$ steps, conditioned on $M_A$.  Finally, we will use the union bound to show that this result holds simultaneously for all choices of $A$ and $M_A$.

To simplify the notation, let us define $h := H_2(S,T|M_A)$, and $p_{st} := \Pr[S=s,T=t|M_A]$; so we have 
\begin{equation}
\sum_{s,t\in\set{0,1}^k} p_{st}^2 \leq 2^{-h}.  
\end{equation}
Note that $h$ and $p_{st}$ depend only on the qubits in the set $A$; so they only depend on those random variables $C(s)_a$ and $D(t)_a$ with $a\in A$.  As shown above, with probability $\geq 1 - e^{-\Omega(n)}$ (over this subset of the random variables $C$ and $D$), 
\begin{equation}
h \geq 2k - m\lg(\tfrac{3}{2}) - \lg(O(n\lg n)).
\end{equation}

We will look at the next $\mya$ steps of the adversary, and we set 
\begin{equation}\label{eq-shrimp2}
\mya := \lfloor h-k \rfloor \geq k - m\lg(\tfrac{3}{2}) - \lg(O(n\lg n)).  
\end{equation}
More precisely, we let $\MMya$ be the set of all possible measurement strategies that an adversary in $\MM'$ may follow for the next $\mya$ steps, after having received measurement outcome $M_A$ on the first $m$ steps.  We let $\Zya := (Z_{m+1},\ldots,Z_{m+\mya})$ be the output of the adversary on the next $\mya$ steps.  Note that this depends only on the qubits outside the set $A$; so it only depends on those random variables $C(s)_a$ and $D(t)_a$ with $a\notin A$.  We refer to this subset of random variables as $\Cya$ and $\Dya$.  We show the following lemma:

\begin{lemma}
Fix a particular subset of qubits $A$ and a particular measurement outcome $M_A$, as described above.  Let $t_0>0$ and $u \geq 1$.  With probability $\geq 1 - \exp(-2t_0^2) - 2\cdot 2^{-u^2}$ (over the choice of $\Cya$ and $\Dya$), the following statement holds:
\begin{equation}
\forall \calM \in \MMya, \; I(\Zya;S,T|M_A) < \frac{\mya}{2} + \frac{t_0\mya}{2^{\mya/2}} + u\cdot O(\sqrt{\log n}).
\end{equation}
\end{lemma}

\noindent
Proof:  We want to upper-bound the quantity 
\begin{equation}
I(\Zya;S,T|M_A) = H(\Zya|M_A) - H(\Zya|S,T,M_A).
\end{equation}
We know that $H(\Zya|M_A) \leq \mya$, since the adversary uses 2-outcome measurements.  We now want to lower-bound $H(\Zya|S,T,M_A)$.  For any $\calM \in \MMya$, we define 
\begin{equation}
Q_\calM := H(\Zya|S,T,M_A), 
\end{equation}
which is a random variable depending on $\Cya$ and $\Dya$.  Note that we can write 
\begin{equation}
Q_\calM = \sum_{s,t\in\set{0,1}^k} p_{st} H(\Zya|S=s,T=t,M_A).
\end{equation}
Let $\mu_\calM := \EE_{\Cya\Dya} Q_\calM$; we will prove a lower bound for $\mu_\calM$ below.  We will then use entropy chaining (Lemma \ref{lem-dudley}) to lower-bound the quantity
\begin{equation}
\inf_{\calM \in \MMya} (Q_\calM - \mu_\calM).
\end{equation}

First, we evaluate $\mu_\calM$.  Using the same argument as in the proof of Theorem \ref{thm-willow}, we get that 
\begin{equation}
\mu_\calM \geq \mya/2.
\end{equation}

We now show several technical facts which are needed in order to apply the entropy chaining argument (Lemma \ref{lem-dudley}).  First, fix some particular strategy $\calM_0 \in \MMya$.  We will show that $Q_{\calM_0}$ is tightly concentrated around its mean.  Observe that $Q_{\calM_0}$ is a sum of $4^k$ random variables, and recall that their dependency graph $\Gamma$ has chromatic number $\chi(\Gamma) \leq 2^k$.  Using Hoeffding's inequality for locally dependent random variables \cite{janson, dubhashi}, we get that 

\begin{equation}
\Pr_{\Cya\Dya}[Q_{\calM_0} < \mu_{\calM_0} - t] \leq \exp\Bigl( -\frac{2t^2}{2^k \sum_{st} (p_{st}\mya)^2} \Bigr)
= \exp\Bigl( -\frac{2t^2 \cdot 2^{\mya}}{\mya^2} \Bigr) \quad (\forall t>0),
\end{equation}
or equivalently
\begin{equation}
\Pr_{\Cya\Dya}\Bigl[Q_{\calM_0} < \mu_{\calM_0} - \frac{t_0 \mya}{2^{\mya/2}}\Bigr] \leq \exp(-2t_0^2), \quad (\forall t_0>0).
\end{equation}

Next, we show that when two strategies $\calM$ and $\calM'$ are ``similar,'' the random variables $Q_\calM$ and $Q_{\calM'}$ are positively correlated.  In particular, suppose that $\calM$ and $\calM'$ behave identically for the first $\ell$ steps.  Let $\Zya_{1,\ldots,\mya}$ and $\Zya'_{1,\ldots,\mya}$ be the output of these two strategies; then $(\Zya_{1,\ldots,\ell},U)$ and $(\Zya'_{1,\ldots,\ell},U)$ have the same distribution.  So we can write
\begin{equation}
\begin{split}
Q_\calM - Q_{\calM'}
 &= H(\Zya_{\ell+1,\ldots,\mya}|\Zya_{1,\ldots,\ell},S,T,M_A) - H(\Zya'_{\ell+1,\ldots,\mya}|\Zya'_{1,\ldots,\ell},S,T,M_A) \\
 &= \sum_{s,t\in\set{0,1}^k} p_{st} [ H(\Zya_{\ell+1,\ldots,\mya}|\Zya_{1,\ldots,\ell},S=s,T=t,M_A) - H(\Zya'_{\ell+1,\ldots,\mya}|\Zya'_{1,\ldots,\ell},S=s,T=t,M_A) ], 
\end{split}
\end{equation}
which again is a sum of $4^k$ locally-dependent random variables.  By Hoeffding's inequality (with local dependencies), 
\begin{equation}
\Pr_{\Cya\Dya}[Q_\calM - Q_{\calM'} - \mu_\calM + \mu_{\calM'} \geq t]
 \leq \exp\Bigl( -\frac{2t^2}{2^k \sum_{st} (p_{st} \cdot 2(\mya-\ell))^2} \Bigr)
 = \exp\Bigl( -\frac{t^2 \cdot 2^{\mya}}{2 (\mya-\ell)^2} \Bigr).
\end{equation}
We can rewrite this bound in terms of a metric $d$ that measures the ``distance'' between strategies.  We define $d$ as follows:
\footnote{Note that $d$ is indeed a metric:  It is easy to see that $d(\calM,\calM') \geq 0$, with equality iff $\calM = \calM'$.  Also, clearly $d(\calM,\calM') = d(\calM',\calM)$.  It remains to show that $d(\calM,\calM'') \leq d(\calM,\calM') + d(\calM',\calM'')$.  We consider two cases.  On one hand, if $\calM'$ satisfies $\ell(\calM,\calM') \leq \ell(\calM,\calM'')$ or $\ell(\calM',\calM'') \leq \ell(\calM,\calM'')$, then the claim follows immediately.  On the other hand, if $\calM'$ satisfies $\ell(\calM,\calM') > \ell(\calM,\calM'')$ and $\ell(\calM',\calM'') > \ell(\calM,\calM'')$, then this is impossible, since $\calM$ and $\calM''$ do not agree at step $\ell(\calM,\calM'')+1$; hence this case cannot occur.}
\begin{equation}
d(\calM,\calM') := \sqrt{2} \cdot 2^{-\mya/2} (\mya-\ell(\calM,\calM')), \text{ where}
\end{equation}
\begin{equation}
\ell(\calM,\calM') := \max \set{\ell \;|\; \text{$0\leq\ell\leq\mya$, and $\calM$ and $\calM'$ behave identically on steps $1,2,\ldots,\ell$}}.
\end{equation}
We then have 
\begin{equation}
\Pr_{\Cya\Dya}[Q_\calM - Q_{\calM'} - \mu_\calM + \mu_{\calM'} \geq t]
 \leq \exp\Bigl( -\frac{t^2}{d(\calM,\calM')^2} \Bigr).
\end{equation}

Next, we bound the covering numbers of $\MMya$ with respect to the metric $d$, and we bound the integral appearing in Lemma \ref{lem-dudley} (``Dudley's entropy integral'').  Using the same argument as in the proof of Theorem \ref{thm-willow}, we get that
\begin{equation}
S \leq O(\sqrt{\log n}).
\end{equation}

Finally, using Lemma \ref{lem-dudley}, we have that:  for all $t_0>0$ and $u \geq 1$, with probability $\geq 1 - \exp(-2t_0^2) - 2\cdot 2^{-u^2}$, the following holds:
\begin{equation}
\forall \calM \in \MMya, \; Q_{\calM} - \mu_{\calM} > - \frac{t_0\mya}{2^{\mya/2}} - uS.
\end{equation}
This implies
\begin{equation}
\forall \calM \in \MMya, \; Q_{\calM} > \frac{\mya}{2} - \frac{t_0\mya}{2^{\mya/2}} - u\cdot O(\sqrt{\log n}).
\end{equation}
Hence
\begin{equation}
\forall \calM \in \MMya, \; I(\Zya;S,T|M_A) < \frac{\mya}{2} + \frac{t_0\mya}{2^{\mya/2}} + u\cdot O(\sqrt{\log n}).
\end{equation}
This proves the claim.  $\square$

{\vskip 11pt}

We now return to the proof of Theorem \ref{thm-squid}.  We take the union bound over all subsets $A\subset[n]$ of size $\abs{A}=m$, and all measurement outcomes $M_A$ that correspond to measurements chosen from the set $L$ and performed on the qubits in the set $A$.  Then, with probability $\geq 1 - 2^{O(n\lg n)} \cdot \exp(-2t_0^2) - 2^{O(n\lg n)} \cdot 2\cdot 2^{-u^2}$, we have that: 
\begin{equation}
\begin{split}
&\text{for all subsets $A$ of size $m$, and all measurement outcomes $M_A$, } \\
&\quad I(\Zya;S,T|M_A) < \frac{\mya}{2} + \frac{t_0\mya}{2^{\mya/2}} + u\cdot O(\sqrt{\log n}).
\end{split}
\end{equation}
By setting $t_0 := \Theta(\sqrt{n\log n})$ and $u := \Theta(\sqrt{n\log n})$, we can make the failure probability exponentially small in $n$.

Hence, for any measurement strategy $\calM \in \MM'$, with output $Z$, and any sequence of measurement outcomes $z_{1,\ldots,m}$, we have
\begin{equation}
I(Z_{m+1,\ldots,m+\mya}; S,T | Z_{1,\ldots,m}=z_{1,\ldots,m}) < \frac{\mya}{2} + O(\sqrt{n} \log n),
\end{equation}
and hence
\begin{equation}\label{eq-shrimp3}
I(Z_{m+1,\ldots,m+\mya}; S,T | Z_{1,\ldots,m}) < \frac{\mya}{2} + O(\sqrt{n} \log n).
\end{equation}

Finally, we consider the remaining steps of the adversary.  Using Holevo's inequality \cite{NC} (see the proof of Theorem \ref{thm-flounder}), we get that
\begin{equation}\label{eq-shrimp4}
I(Z_{m+\mya+1,\ldots,n}; S,T | Z_{1,\ldots,m+\mya}) \leq n-m-\mya.
\end{equation}

Combining equations (\ref{eq-shrimp1}), (\ref{eq-shrimp3}) and (\ref{eq-shrimp4}), we get that
\begin{equation}
\begin{split}
I(Z;S,T) &\leq m\lg(\tfrac{3}{2}) + \lg(O(n\lg n)) + \tfrac{1}{2}\mya + O(\sqrt{n} \log n) + n-m-\mya \\
&= n - m\lg(\tfrac{4}{3}) - \tfrac{1}{2}\mya + O(\sqrt{n} \log n).
\end{split}
\end{equation}
From equations (\ref{eq-penguin}), (\ref{eq-shrimp0}) and (\ref{eq-shrimp2}), we have that 
\begin{equation}
n \leq \tfrac{k}{1-h(p_e)} + O(\sqrt{n}) \approx (2.5056)k + O(\sqrt{n}),
\end{equation}
\begin{equation}
m = \lfloor \tfrac{k}{\lg(8/3)} \rfloor \approx \lfloor (0.7067)k \rfloor, 
\end{equation}
\begin{equation}
\begin{split}
\mya &\geq k - m\lg(\tfrac{3}{2}) -\lg(O(n\lg n)) \\
 &\geq k - \tfrac{k\lg(3/2)}{\lg(8/3)} -\lg(O(n\lg n)) \\
 &\approx (0.5866)k -\lg(O(n\lg n)).
\end{split}
\end{equation}
Combining these bounds, we get that 
\begin{equation}
I(Z;S,T) \leq (1.9190)k + O(\sqrt{n} \log n),
\end{equation}
as desired.  $\square$


\subsection*{Acknowledgements}

It is a pleasure to thank Anne Broadbent, Daniel Gottesman, Jonathan Katz, Dianne O'Leary, Rene Peralta, Christian Schaffner, Jake Taylor, and Stephanie Wehner, for helpful suggestions about this work.  This work is a contribution of NIST, an agency of the US government, and is not subject to US copyright laws.


\appendix

\section{Facts about LOCC measurement strategies}
\label{app-locc-epsnet}

\begin{lemma}(restatement of Lemma \ref{lem-locc-rank1})
Let $\calM$ be any 1-pass LOCC strategy in the isolated qubits model, which uses $q$-outcome measurements and returns output $Z$.  Then there exists $\calM'$, a 1-pass LOCC strategy in the isolated qubits model, which uses $q$-outcome measurements and returns output $Z'$, and has the following additional properties:  
\begin{enumerate}
\item $I(Z';U) \geq I(Z;U)$ (when playing the state discrimination game shown above).
\item In every measurement performed by $\calM'$, the POVM elements all have rank 1.
\end{enumerate}
\end{lemma}

\noindent
Proof:  We will construct the strategy $\calM'$ as follows.  Consider what the strategy $\calM$ does at step $a$, given some prior history $z_{<a}$.  Any POVM element $M_a(z_{<a},\zeta)$ that has rank $>1$ can be written in the form $\alpha I + \beta \ket{\varphi}\bra{\varphi}$, where $\alpha>0$, $\beta \geq 0$.  We now construct a new POVM measurement, by replacing $M_a(z_{<a},\zeta)$ with two operators $\alpha I$ and $\beta \ket{\varphi}\bra{\varphi}$.  This new measurement can simulate the original measurement, by identifying the measurement outcomes $\alpha I$ and $\beta \ket{\varphi}\bra{\varphi}$ with the original measurement outcome $\zeta$.

In this way, one can replace each measurement in $\calM$ with a measurement that consists of at most $q$ POVM elements that have rank 1, and at most $q$ POVM elements that are multiples of $I$.  This strategy is equivalent to a probabilistic mixture of strategies, where each strategy uses measurements with at most $q$ POVM elements, each of which has rank 1.  By convexity of the mutual information $I(Z;U)$ (as a function of the conditional distribution $\Pr[Z=z|U=u]$, keeping the marginal distribution $\Pr[U=u]$ fixed), there must be a pure strategy $\calM'$ that achieves $I(Z';U) \geq I(Z;U)$, and uses measurements with at most $q$ POVM elements, each of which has rank 1.  $\square$

\begin{lemma}(restatement of Lemma \ref{lem-epsnet-2})
Let $q = 2$.  For any $0 < \eps \leq 1$, there exists an $\eps$-net $L$ for $\calS$, with respect to the metric $t$, that has cardinality $|L| \leq C/\eps^2$ (where $C$ is some numerical constant).  Equivalently, we have $N(\calS,t,\eps) \leq C/\eps^2$.
\end{lemma}

\noindent
Proof:  When $q = 2$, we can write the set $\calS$ and the metric $t$ in a simpler form:
\begin{equation}
\calS = \set{(\ket{\varphi}\bra{\varphi}, I-\ket{\varphi}\bra{\varphi}) \text{ s.t. } \ket{\varphi} \in \CC^2, \; \braket{\varphi}{\varphi} = 1},
\end{equation}
\begin{equation}
t\bigl((\ket{\varphi}\bra{\varphi}, I-\ket{\varphi}\bra{\varphi}), (\ket{\theta}\bra{\theta}, I-\ket{\theta}\bra{\theta})\bigr) = \norm{\ket{\varphi}\bra{\varphi} - \ket{\theta}\bra{\theta}}.
\end{equation}
Let $\calB := \set{\ket{\varphi}\bra{\varphi} \text{ s.t. } \ket{\varphi} \in \CC^2, \; \braket{\varphi}{\varphi} = 1}$, and note that 
\begin{equation}
N(\calS,t,\eps) \leq N(\calB,\norm{\cdot},\eps).
\end{equation}
It follows from standard arguments 
\footnote{Let $\calB_2 = \set{\ket{\varphi} \in \CC^2, \; \braket{\varphi}{\varphi} = 1}$.  We claim that, for all $\eps \leq 1$, $N(\calB, \norm{\cdot}, \eps) \leq N(\calB_2, \norm{\cdot}_2, \eps/3)$.  
This follows because, for any $\ket{\varphi}\bra{\varphi}, \ket{\varphi'}\bra{\varphi'} \in \calB$, such that $\ket{u} := \ket{\varphi'} - \ket{\varphi}$ satisfies $\norm{u}_2 \leq 1$, we can write $\norm{\ket{\varphi}\bra{\varphi} - \ket{\varphi'}\bra{\varphi'}} = \norm{ -\ket{u}\bra{\varphi} - \ket{\varphi}\bra{u} - \ket{u}\bra{u}} \leq 2\norm{u}_2 + \norm{u}_2^2 \leq 3\norm{u}_2$.  
Finally, it is easy to see that $N(\calB_2, \norm{\cdot}_2, \eps/3) \leq O(1/\eps^2)$.}
that $N(\calB, \norm{\cdot}, \eps) \leq O(1/\eps^2)$.  $\square$

\begin{lemma}(restatement of Lemma \ref{lem-epsnet-q})
Let $q \geq 2$.  For any $0 < \eps \leq 1$, there exists an $\eps$-net $L$ for $\calS$, with respect to the metric $t$, that has cardinality $|L| \leq (C/\eps)^{3q}$ (where $C$ is some numerical constant).  Equivalently, we have $N(\calS,t,\eps) \leq (C/\eps)^{3q}$.
\end{lemma}

\noindent
Proof:  Observe that $\calS \subset (\calS'_1)^q$, where $\calS'_1 := \set{M \in \CC^{2\times 2} \;|\; 0 \preceq M \preceq I, \; \rank(M) = 1}$.  This implies:
\footnote{This follows because, given an $(\eps/2)$-net for $\calS'_1$, we can take its $q$-fold Cartesian product, ``round'' each point to the nearest point in $\calS$, and get an $\eps$-net for $\calS$.}
\begin{equation}
N(\calS,t,\eps) \leq N(\calS'_1, \norm{\cdot}, \eps/2)^q.
\end{equation}

Next, let $\calB := \set{\ket{\varphi}\bra{\varphi} \text{ s.t. } \ket{\varphi} \in \CC^2, \; \braket{\varphi}{\varphi} = 1}$.  Note that we can write $\calS'_1 = \set{\lambda M \;|\; \lambda \in [0,1], \; M \in \calB}$.  This implies:
\footnote{To see this, let $E_1$ be any $(\eps/4)$-net for $[0,1]$, and let $E_2$ be any $(\eps/4)$-net for $\calB$.  We claim that $F := \set{\tilde{\lambda} \tilde{M} \;|\; \tilde{\lambda} \in E_1, \; \tilde{M} \in E_2}$ is an $(\eps/2)$-net for $\calS'_1$.  To see this, let $\lambda M$ be any element of $\calS'_1$.  Then there exists some $\tilde{\lambda} \tilde{M} \in F$, such that $\norm{\lambda M - \tilde{\lambda} \tilde{M}} \leq \norm{\lambda M - \tilde{\lambda} M} + \norm{\tilde{\lambda} M - \tilde{\lambda} \tilde{M}} \leq \abs{\lambda-\tilde{\lambda}} + \norm{M-\tilde{M}} \leq \eps/4 + \eps/4 = \eps/2$.}
\begin{equation}
N(\calS'_1, \norm{\cdot}, \eps/2) \leq N([0,1], \abs{\cdot}, \eps/4) N(\calB, \norm{\cdot}, \eps/4).
\end{equation}

It is easy to see that $N([0,1], \abs{\cdot}, \eps/4) \leq O(1/\eps)$, and it follows from standard arguments 
\footnote{Let $\calB_2 = \set{\ket{\varphi} \in \CC^2, \; \braket{\varphi}{\varphi} = 1}$.  We claim that, for all $\eps \leq 1$, $N(\calB, \norm{\cdot}, \eps/4) \leq N(\calB_2, \norm{\cdot}_2, \eps/12)$.  
This follows because, for any $\ket{\varphi}\bra{\varphi}, \ket{\varphi'}\bra{\varphi'} \in \calB$, such that $\ket{u} := \ket{\varphi'} - \ket{\varphi}$ satisfies $\norm{u}_2 \leq 1$, we can write $\norm{\ket{\varphi}\bra{\varphi} - \ket{\varphi'}\bra{\varphi'}} = \norm{ -\ket{u}\bra{\varphi} - \ket{\varphi}\bra{u} - \ket{u}\bra{u}} \leq 2\norm{u}_2 + \norm{u}_2^2 \leq 3\norm{u}_2$.  
Finally, it is easy to see that $N(\calB_2, \norm{\cdot}_2, \eps/12) \leq O(1/\eps^2)$.}
that $N(\calB, \norm{\cdot}, \eps/4) \leq O(1/\eps^2)$.  $\square$

\begin{lemma}[restatement of Lemma \ref{lem-locc-epsnet}]
Let $\calM$ be any 1-pass LOCC strategy in the isolated qubits model, which uses $q$-outcome measurements, where all POVM elements have rank 1, and which has output $Z$.  Fix some $0 < \eps \leq 1/(qne)$, and let $L$ be the $\eps$-net for $\calS$ defined above.  Let $\calM'$ be the strategy that is obtained by duplicating the strategy $\calM$, and replacing each measurement $M \in \calS$ with the best approximating measurement $\tilde{M} \in L$.  Let $Z'$ be the output of the strategy $\calM'$.  Then 
\begin{equation}
\abs{I(Z';U) - I(Z;U)} \leq 2qn^2\eps + 2\eta(qn\eps), 
\end{equation}
where $\eta(x) := -x\lg x$.
\end{lemma}

\noindent
Proof:  For any $u \in \set{0,1}^{\nya}$, let $Z|_{U=u}$ be the random variable $Z$ conditioned on the event $U=u$; define $Z'|_{U=u}$ similarly.  We will show that $Z|_{U=u}$ and $Z'|_{U=u}$ have nearly the same distribution, compared using total variation distance (denoted $\Delta(\cdot,\cdot)$).  To see this, let us define a sequence of strategies that interpolate between $\calM$ and $\calM'$.  For $a=0,1,2,\ldots,n$, we define a strategy $\calM^{(a)}$ (whose output is denoted $Z^{(a)}$) that does the same measurements as $\calM'$ for steps $1,2,\ldots,a$, and does the same measurements as $\calM$ for steps $a+1,a+2,\ldots,n$.  Note that $\calM^{(0)} = \calM$ and $\calM^{(n)} = \calM'$, and we have
\begin{equation}
\Delta(Z|_{U=u}, Z'|_{U=u}) \leq \sum_{a=0}^{n-1} \Delta(Z^{(a)}|_{U=u}, Z^{(a+1)}|_{U=u}).
\end{equation}

We now want to bound 
\begin{equation}
\Delta(Z^{(a)}|_{U=u}, Z^{(a+1)}|_{U=u}) = \sum_{z \in [q]^n} \abs{\Pr[Z^{(a)}=z|U=u] - \Pr[Z^{(a+1)}=z|U=u]}.
\end{equation}
The state of the $n$ qubits is given by $\ket{E(u)} = \Tensor_{a=1}^n \ket{\alpha_{E(u)_a}}$; to simplify notation, let us call this state $\ket{\psi} = \Tensor_{a=1}^n \ket{\psi_a}$.  We will use the following notation:  the strategy $\calM$ is described by POVM elements $M_i(z_{\leq i})$, with the choice of which qubit to measure next being specified by $Q_i(z_{<i})$; the strategy $\calM'$ is described by slightly different POVM elements $M'_i(z_{\leq i})$, and the same qubit choices $Q_i(z_{<i})$.  Then we can write 
\begin{equation}
\Pr[Z^{(a)}=z|U=u] = \prod_{i=1}^a \bra{\psi_{Q_i(z_{<i})}} M'_i(z_{\leq i}) \ket{\psi_{Q_i(z_{<i})}} \cdot 
\prod_{i=a+1}^n \bra{\psi_{Q_i(z_{<i})}} M_i(z_{\leq i}) \ket{\psi_{Q_i(z_{<i})}}.
\end{equation}
Hence
\begin{equation}
\begin{split}
\Delta(Z^{(a)}|_{U=u}, Z^{(a+1)}|_{U=u})
 = \sum_{z \in [q]^n} 
&\prod_{i=1}^a \bra{\psi_{Q_i(z_{<i})}} M'_i(z_{\leq i}) \ket{\psi_{Q_i(z_{<i})}} \cdot \\
&\Bigl\lvert \bra{\psi_{Q_{a+1}(z_{<a+1})}} \bigl[ M_{a+1}(z_{\leq a+1}) - M'_{a+1}(z_{\leq a+1}) \bigr] \ket{\psi_{Q_{a+1}(z_{<a+1})}} \Bigr\rvert \cdot \\
&\prod_{i=a+2}^n \bra{\psi_{Q_i(z_{<i})}} M_i(z_{\leq i}) \ket{\psi_{Q_i(z_{<i})}}
\end{split}
\end{equation}
Now we can use the bound $\norm{M_{a+1}(z_{\leq a+1}) - M'_{a+1}(z_{\leq a+1})} \leq \eps$, and we can evaluate the sum over $z$, using the fact that for any $z_{<i}$, $\sum_{z_i} M_i(z_{\leq i}) = I$ (and similarly for $M'_i(z_{\leq i})$).  We get that
\begin{equation}
\Delta(Z^{(a)}|_{U=u}, Z^{(a+1)}|_{U=u}) \leq q\eps,
\end{equation}
and therefore
\begin{equation}
\Delta(Z|_{U=u}, Z'|_{U=u}) \leq qn\eps,
\end{equation}
which shows that $Z|_{U=u}$ and $Z'|_{U=u}$ have nearly identical distributions, as desired.

In the remainder of the proof, we will bound the difference between $I(Z;U)$ and $I(Z';U)$.  First, using (the classical case of) Fannes' inequality \cite{NC}, and assuming $qn\eps \leq 1/e$, we get that 
\begin{equation}
\abs{H(Z|U=u) - H(Z'|U=u)} \leq qn^2\eps + \eta(qn\eps), 
\end{equation}
where $\eta(x) := -x\lg x$.  This implies 
\begin{equation}
\abs{H(Z|U) - H(Z'|U)}
 = \bigl\lvert 2^{-\nya} \sum_{u\in\set{0,1}^{\nya}} (H(Z|U=u) - H(Z'|U=u)) \bigr\rvert
 \leq qn^2\eps + \eta(qn\eps).
\end{equation}
Next, we can bound the total variation distance between $Z$ and $Z'$ as follows:
\begin{equation}
\begin{split}
\Delta(Z,Z') 
 &= \sum_{z\in[q]^n} \bigl\lvert 2^{-\nya} \sum_{u\in\set{0,1}^{\nya}} (\Pr[Z=z|U=u] - \Pr[Z'=z|U=u]) \bigr\rvert \\
 &\leq 2^{-\nya} \sum_{u\in\set{0,1}^{\nya}} \Delta(Z|_{U=u}, Z'|_{U=u}) \\
 &\leq qn\eps.
\end{split}
\end{equation}
Then, by Fannes' inequality, 
\begin{equation}
\abs{H(Z) - H(Z')} \leq qn^2\eps + \eta(qn\eps).
\end{equation}
Combining these bounds, we get
\begin{equation}
\abs{I(Z;U) - I(Z';U)} 
\leq 2qn^2\eps + 2\eta(qn\eps),
\end{equation}
as desired.  $\square$


\section{High success probability implies high mutual information}
\label{app-ostrich}

\begin{lemma}[restatement of Lemma \ref{lem-ostrich}]
Suppose that $\Pr[Z=U] \geq 1-\eps$, and $\eps$ is sufficiently small that $2\sqrt{\eps} + 2^{-\nya} \leq 1/e$.  Then $I(Z;U) \geq (1-5\sqrt{\eps}) \nya - \eta(2\sqrt{\eps})$, where $\eta(x) := -x \lg x$.
\end{lemma}

\noindent
Proof:  First, we claim that, for most $u\in\set{0,1}^{\nya}$, $\Pr[Z=u|U=u]$ is close to 1.  To see this, suppose $u$ is chosen uniformly at random in $\set{0,1}^{\nya}$, and define $\gamma(u) := 1-\Pr[Z=u|U=u]$.  Note that $\gamma(u) \geq 0$ and 
\begin{equation}
\EE_u[\gamma(u)] = 2^{-\nya} \sum_{u\in\set{0,1}^{\nya}} (1-\Pr[Z=u|U=u]) = 1-\Pr[Z=U] \leq \eps.  
\end{equation}
By Markov's inequality, for any $C \geq 1$, $\Pr_u[\gamma(u) \geq C\eps] \leq \tfrac{1}{C}$.  Therefore, there exists a subset $S \subseteq \set{0,1}^{\nya}$ of size $\abs{S} \geq 2^{\nya} (1-\tfrac{1}{C})$, such that for all $u \in S$, 
\begin{equation}
\Pr[Z=u|U=u] > 1-C\eps.  
\end{equation}
We need to choose $C$ such that both $\tfrac{1}{C}$ and $C\eps$ are small.  For concreteness, we set $C = \tfrac{1}{\sqrt{\eps}}$, which implies that $\tfrac{1}{C} = \sqrt{\eps} = C\eps$.

We now show that $H(Z|U)$ is small.  We write $H(Z|U) = 2^{-\nya} \sum_{u\in\set{0,1}^{\nya}} H(Z|U=u)$, and we upper-bound $H(Z|U=u)$.  First, consider the case where $u \in S$.  We bound the total-variation distance between the random variables $Z|_{U=u}$ and $U|_{U=u}$ as follows:  (note that $U|_{U=u}$ equals $u$ with probability 1)
\begin{equation}
\begin{split}
\Delta(Z|_{U=u},U|_{U=u}) &= \abs{\Pr[Z=u|U=u]-1} + \sum_{z\neq u} \abs{\Pr[Z=z|U=u]-0} \\
 &= 1 - \Pr[Z=u|U=u] + \Pr[Z\neq u|U=u] \\
 &= 2(1-\Pr[Z=u|U=u]) \\
 &< 2C\eps = 2\sqrt{\eps}.
\end{split}
\end{equation}
Using Fannes' inequality \cite{NC} (note that $2\sqrt{\eps} \leq 1/e$), we get that
\begin{equation}
H(Z|U=u) = \abs{H(Z|U=u) - H(U|U=u)} \leq 2\sqrt{\eps} \cdot \nya + \eta(2\sqrt{\eps}).  
\end{equation}
Next, consider the case where $u \notin S$.  Here we use the trivial bound, $H(Z|U=u) \leq \nya$.  We now bound $H(Z|U)$ as follows:
\begin{equation}
H(Z|U) \leq 2^{-\nya} \abs{S} (2\sqrt{\eps} \cdot \nya + \eta(2\sqrt{\eps})) + 2^{-\nya} \abs{S^c} \nya.
\end{equation}
The right-hand side is largest when $\abs{S} = 2^{\nya}(1-\sqrt{\eps})$, so we get
\begin{equation}\label{eq-minnow-1}
\begin{split}
H(Z|U) &\leq (1-\sqrt{\eps}) (2\sqrt{\eps} \cdot \nya + \eta(2\sqrt{\eps})) + \sqrt{\eps} \cdot \nya \\
 &< 3\sqrt{\eps} \cdot \nya + \eta(2\sqrt{\eps}).
\end{split}
\end{equation}

Next, we observe that, for most $z \in \set{0,1}^{\nya}$, $\Pr[Z=z]$ is not much smaller than $2^{-\nya}$.  More precisely, for all $z \in S$, we have a lower bound:
\begin{equation}
\Pr[Z=z] \geq 2^{-\nya} \Pr[Z=z|U=z] > 2^{-\nya} (1-\sqrt{\eps}).
\end{equation}
We also show a (loose) upper-bound on $\Pr[Z=z]$, when $z \in S$, as follows:
\begin{equation}
\begin{split}
\Pr[Z=z] &= 2^{-\nya} \sum_{u \in \set{0,1}^{\nya}} \Pr[Z=z|U=u] \\
 &\leq 2^{-\nya} \abs{S\setminus\set{z}} \sqrt{\eps} + 2^{-\nya} + 2^{-\nya} \abs{S^c}
\end{split}
\end{equation}
The right-hand side is largest when $\abs{S} = 2^{\nya}(1-\sqrt{\eps})$, so we get
\begin{equation}
\begin{split}
\Pr[Z=z] &< (1-\sqrt{\eps}) \sqrt{\eps} + 2^{-\nya} + \sqrt{\eps} \\
 &< 2\sqrt{\eps} + 2^{-\nya}.
\end{split}
\end{equation}

Finally, we will show that $H(Z)$ is large.  First, we write 
\begin{equation}
H(Z) \geq \sum_{z\in S} \eta(\Pr[Z=z]).  
\end{equation}
Note that $\eta(x) := -x \lg x$ is increasing on the interval $[0,1/e]$.  From the previous paragraph, we know that for all $z \in S$, we have $2^{-\nya} (1-\sqrt{\eps}) < \Pr[Z=z] < 2\sqrt{\eps} + 2^{-\nya} < 1/e$.  So we can write
\begin{equation}\label{eq-minnow-2}
\begin{split}
H(Z) &\geq \sum_{z\in S} \eta(2^{-\nya} (1-\sqrt{\eps})) \\
 &= \abs{S} 2^{-\nya} (1-\sqrt{\eps}) (-1) \lg(2^{-\nya} (1-\sqrt{\eps})) \\
 &\geq (1-\sqrt{\eps})^2 (\nya - \lg(1-\sqrt{\eps})) \\
 &> (1-2\sqrt{\eps}) \nya.
\end{split}
\end{equation}

Finally, we combine equations (\ref{eq-minnow-1}) and (\ref{eq-minnow-2}) to get the desired lower bound on $I(Z;U) = H(Z) - H(Z|U)$.  $\square$


\section{Entropy chaining}
\label{app-entropy-chaining}

We prove a variant of Dudley's inequality, for bounding the expected supremum of a family of correlated random variables, $\EE \sup_t X_t$, using entropy chaining.  Our claim is a slight generalization of the usual statement of Dudley's inequality, in that it allows the random variables $X_t$ to have different means; also, we state our result as a tail bound on $\sup_t X_t$, which is stronger than the usual form of Dudley's inequality.  Nonetheless, the proof is more or less the same as the usual one; see, e.g., \cite{talagrand}.

\begin{lemma}[Dudley's inequality tail bound]\label{lem-dudley}
Let $\set{X_t \;|\; t\in T}$ be a family of random variables taking values in $\RR$.  Define $\mu_t := \EE X_t$.  

Let $d(\cdot,\cdot)$ be a metric on the set $T$, such that the following ``increment condition'' holds:  
\begin{equation}\label{dudley-increment}
\Pr[X_s - X_t - \mu_s + \mu_t > u] \leq \exp(-u^2 / d(s,t)^2), \quad \forall s,t\in T, \quad \forall u>0.
\end{equation}
(Note that, by exchanging $s$ and $t$, this also implies a similar bound on the lower tail of $X_s - X_t$.)  Also, suppose that, for any sequence $s_1,s_2,s_3,\ldots \in T$, and any $t \in T$, if $\lim_{j\rightarrow\infty} d(s_j,t) = 0$, then $\lim_{j\rightarrow\infty} \mu_{s_j} = \mu_t$.

Suppose there exist $t_0 \in T$ and $\delta,\eps > 0$, such that $Pr[X_{t_0} - \mu_{t_0} > \delta] \leq \eps$.  Then we have the following bound:
\begin{equation}\label{dudley-upper-tail}
\Pr[\sup_t (X_t - \mu_t) > \delta + uS] \leq \eps + 2\cdot 2^{-u^2}, \quad \forall u\geq 1, 
\end{equation}
where 
\begin{equation}\label{dudley-S}
S \leq C_0 \int_0^\infty \sqrt{\log N(T,d,\eps)} d\eps, 
\end{equation}
$C_0$ is a numerical constant, and $N(T,d,\eps)$ is the covering number, i.e., the minimum cardinality of an $\eps$-net for the set $T$ with respect to the metric $d(\cdot,\cdot)$.

By applying the same argument to the random variables $\set{-X_t \;|\; t\in T}$, we also have a lower bound.  Suppose there exist $t_0 \in T$ and $\delta,\eps > 0$, such that $Pr[X_{t_0} - \mu_{t_0} < -\delta] \leq \eps$.  Then:
\begin{equation}\label{dudley-lower-tail}
\Pr[\inf_t (X_t - \mu_t) < -\delta - uS] \leq \eps + 2\cdot 2^{-u^2}, \quad \forall u\geq 1.
\end{equation}
\end{lemma}

\noindent
Proof:  We use a standard entropy chaining argument \cite{talagrand}.  Fix some $r \geq 2$, and choose some integer $j_0$ such that $r^{-(j_0+1)} < \text{diam}(T) \leq r^{-j_0}$.  For all $j \geq j_0$, we will construct sets $\Pi_j \subset T$ and maps $\pi_j:\; T \rightarrow \Pi_j$ that have the following properties:  
\begin{equation}\label{dudley-ppty1}
\pi_{j_0}(t) = t_0, \quad \forall t\in T, 
\end{equation}
\begin{equation}\label{dudley-ppty2}
\lim_{j\rightarrow\infty} d(\pi_j(t), t) = 0, \quad \forall t\in T, 
\end{equation}
\begin{equation}\label{dudley-ppty3}
d(\pi_j(t), \pi_{j-1}(t)) \leq 2r^{-(j-1)}, \quad \forall t\in T, \quad \forall j\geq j_0+1.
\end{equation}
(Intuitively, for each $t\in T$, the sequence of points $\set{\pi_j(t) \;|\; j=j_0, j_0+1, j_0+2, \ldots}$ starts at $t_0$ and quickly converges to $t$.)  Also note that equation (\ref{dudley-ppty2}) implies that 
\begin{equation}\label{dudley-ppty4}
\lim_{j\rightarrow\infty} \mu_{\pi_j(t)} = \mu_t, \quad \forall t\in T, 
\end{equation}

We will construct the sets $\Pi_j$ and maps $\pi_j$ later.  In the mean time, note that 
\begin{equation}
X_t - X_{t_0} = \sum_{j\geq j_0+1} X_{\pi_j(t)} - X_{\pi_{j-1}(t)}, 
\end{equation}
\begin{equation}
\mu_t - \mu_{t_0} = \sum_{j\geq j_0+1} \mu_{\pi_j(t)} - \mu_{\pi_{j-1}(t)}.
\end{equation}
Fix any real numbers $a_j > 0$ (for all $j\geq j_0+1$).  (We will choose values for the $a_j$ later.)  Define $S := \sum_{j\geq j_0+1} a_j$, and fix any $u>0$.  Note that, for any $t\in T$, if 
\begin{equation}
X_{\pi_j(t)} - X_{\pi_{j-1}(t)} \leq \mu_{\pi_j(t)} - \mu_{\pi_{j-1}(t)} + ua_j, \quad \forall j\geq j_0+1, 
\end{equation}
then $X_t - X_{t_0} \leq \mu_t - \mu_{t_0} + uS$.  Moreover, using the increment condition (\ref{dudley-increment}), we have that 
\begin{equation}
\Pr[X_{\pi_j(t)} - X_{\pi_{j-1}(t)} > \mu_{\pi_j(t)} - \mu_{\pi_{j-1}(t)} + ua_j]
\leq \exp(-u^2 a_j^2 / (2r^{-(j-1)})^2).
\end{equation}
Hence, using the union bound, we get that 
\begin{equation}
\begin{split}
\Pr[\exists t\in T &\text{ s.t. } X_t - X_{t_0} > \mu_t - \mu_{t_0} + uS] \\
&\leq \Pr[\exists t\in T, \; \exists j\geq j_0+1, \text{ s.t. } X_{\pi_j(t)} - X_{\pi_{j-1}(t)} > \mu_{\pi_j(t)} - \mu_{\pi_{j-1}(t)} + ua_j] \\
&\leq \sum_{j\geq j_0+1} |\Pi_j| |\Pi_{j-1}| \exp(-u^2 a_j^2 / (2r^{-(j-1)})^2).
\end{split}
\end{equation}

Now set $a_j := 2r^{-(j-1)} \sqrt{\log(2^{j-j_0} |\Pi_j| |\Pi_{j-1}|)}$, and assume that $u^2 \geq 1$.  Then we have 
\begin{equation}
\begin{split}
\Pr[\exists t\in T &\text{ s.t. } X_t - X_{t_0} > \mu_t - \mu_{t_0} + uS] \\
&\leq \sum_{j\geq j_0+1} |\Pi_j| |\Pi_{j-1}| (2^{j-j_0} |\Pi_j| |\Pi_{j-1}|)^{-u^2} \\
&\leq \sum_{j\geq j_0+1} 2^{-(j-j_0) u^2} = 2^{-u^2} \sum_{a=0}^\infty 2^{-au^2} \\
&\leq 2^{-u^2} \sum_{a=0}^\infty 2^{-a} < 2\cdot 2^{-u^2}.
\end{split}
\end{equation}
We can rewrite this as $\Pr[\sup_{t\in T} (X_t - \mu_t) > X_{t_0} - \mu_{t_0} + uS] < 2\cdot 2^{-u^2}$.  This now implies the claimed bound (\ref{dudley-upper-tail}); and by applying the same argument to the random variables $\set{-X_t \;|\; t\in T}$, we also get the bound (\ref{dudley-lower-tail}).

It remains to construct the sets $\Pi_j$ and maps $\pi_j$, and prove the upper bound on $S$ shown in (\ref{dudley-S}).  For each $j\geq j_0$, we choose the set $\Pi_j$ to be an $\eps$-net for the set $T$, with $\eps = r^{-j}$, and with respect to the metric $d(\cdot,\cdot)$.  In particular, we choose $\Pi_j$ to be an $\eps$-net of minimum cardinality, so that $|\Pi_j| = N(T,d,r^{-j})$.  For notational convenience, we define $N_j := |\Pi_j|$.  In the case of $j=j_0$, we let $\Pi_{j_0} = \set{t_0}$, recalling that $\text{diam}(T) \leq r^{-j_0}$.  We define $\pi_j$ to be the map that, given any point $t\in T$, returns the nearest point in $\Pi_j$; hence, $d(\pi_j(t), t) \leq r^{-j}$.  Note that equations (\ref{dudley-ppty1}) and (\ref{dudley-ppty2}) are satisfied, and (\ref{dudley-ppty3}) follows from the triangle inequality.  

We upper-bound $S$ as follows:
\begin{equation}
\begin{split}
S &= \sum_{j\geq j_0+1} a_j = \sum_{j\geq j_0+1} 2r^{-(j-1)} \sqrt{\log(2^{j-j_0} N_j N_{j-1})} \\
&\leq \sum_{j\geq j_0+1} 2r^{-(j-1)} \Bigl( \sqrt{(j-j_0)\log2} + \sqrt{\log N_j} + \sqrt{\log N_{j-1}} \Bigr) \\
&\leq r^{-j_0} \sum_{j=1}^\infty 2r^{-(j-1)} \sqrt{j\log2} + (r+1) \sum_{j\geq j_0} 2r^{-j} \sqrt{\log N_j} \\
&= r^{-j_0} K(r) + (r+1) \sum_{j\geq j_0} 2r^{-j} \sqrt{\log N_j}, 
\end{split}
\end{equation}
where we used the fact that $\sqrt{a+b} \leq \sqrt{a} + \sqrt{b}$ (for all $a,b \geq 0$), and we defined 
$K(r) := \sum_{j=1}^\infty 2r^{-(j-1)} \sqrt{j\log2}$.  Next, recall that $\text{diam}(T) > r^{-(j_0+1)} \geq 2r^{-(j_0+2)}$, and hence $N_{j_0+2} \geq 2$.  So we can write
\begin{equation}
S \leq \biggl( \frac{r^2 K(r)}{2\sqrt{\log2}} + r+1 \biggr) \sum_{j\geq j_0} 2r^{-j} \sqrt{\log N_j}.
\end{equation}
We will now replace the sum on the right hand side by an integral.  Note that, for any $\eps \leq r^{-j}$, we have $N(T,d,\eps) \geq N_j$.  So we can write
\begin{equation}
\int_{r^{-(j+1)}}^{r^{-j}} \sqrt{\log N(T,d,\eps)} d\eps
\geq (1-\tfrac{1}{r}) r^{-j} \sqrt{\log N_j},
\end{equation}
and hence
\begin{equation}
S \leq \biggl( \frac{r^2 K(r)}{2\sqrt{\log2}} + r+1 \biggr) \cdot 2 (1-\tfrac{1}{r})^{-1} 
\int_0^{r^{-j_0}} \sqrt{\log N(T,d,\eps)} d\eps.
\end{equation}
Note that we can extend the integral over the interval $[0,\infty)$ without weakening the bound; for when $\eps \geq r^{-j_0}$, we have $N(T,d,\eps) = 1$, hence $\sqrt{\log N(T,d,\eps)} = 0$.  Now set $r\geq 2$ to be some numerical constant.  This proves equation (\ref{dudley-S}).  $\square$


\section{Shannon's noisy coding theorem}
\label{app-noisy-coding}

\begin{proposition}(restatement of Prop. \ref{prop-sea-cucumber})
Fix any constants $\lambda \geq 1$, $0 < \tau \leq \tfrac{1}{2} - p_e \approx 0.3536$, and $\theta > \tau h'(p_e) \approx (2.5431) \tau$.  Then for all sufficiently large $n$, the following statement holds:  with probability $\geq 1-\tfrac{1}{\lambda}$ (over the choice of $C$), we have 
\begin{equation}
\Pr[\Shat=S] \geq 1 - \lambda \bigl[ e^{-2\tau^2 n} + 2^{-n (\theta - \tau h'(p_e))} \bigr]
\geq 1 - \lambda e^{-\Omega(n)}.
\end{equation}
\end{proposition}

\noindent
Proof:  We can view $\Pr[\Shat \neq S]$ as a random variable depending on the choice of the random code $C$.  We then calculate $\EE_C \Pr[\Shat \neq S]$.

We can upper-bound $\Pr[\Shat \neq S]$ as follows:
\begin{equation}
\Pr[\Shat \neq S]
\leq \Pr[d_H(C(S),Z) > r] + \Pr[d_H(C(S),Z) \leq r \text{ and } \exists t\in \set{0,1}^k \text{ s.t. } t\neq S, \; d_H(C(t),Z) \leq r].
\end{equation}
Let $N_e$ be the number of errors introduced by the channel $BSC(p_e)$, acting independently on the $n$ bits of the string $C(S)$.  Then $N_e = d_H(C(S),Z)$, $\EE N_e = np_e$, and by Hoeffding's inequality, $\Pr[N_e > r] \leq e^{-2\tau^2 n}$.  So we have 
\begin{equation}\label{eq-bat}
\begin{split}
\Pr[\Shat \neq S]
&\leq e^{-2\tau^2 n} + \Pr[\exists t\in \set{0,1}^k \text{ s.t. } t\neq S, \; d_H(C(t),Z) \leq r] \\
&= e^{-2\tau^2 n} + 2^{-k} \sum_{s\in\set{0,1}^k} \Pr[\exists t\in \set{0,1}^k \text{ s.t. } t\neq S, \; d_H(C(t),Z) \leq r | S=s] \\
&\leq e^{-2\tau^2 n} + 2^{-k} \sum_{s\in\set{0,1}^k} \sum_{t\in \set{0,1}^k \setminus \set{s}} \Pr[d_H(C(t),Z) \leq r | S=s] \\
&= e^{-2\tau^2 n} + 2^{-k} \sum_{s\in\set{0,1}^k} \sum_{t\in \set{0,1}^k \setminus \set{s}} \sum_{z\in\set{0,1}^n} 1[d_H(C(t),z) \leq r] \Pr[Z=z | S=s]. \\
\end{split}
\end{equation}

We now bound $\EE_C \Pr[\Shat \neq S]$, taking the expectation over the choice of the random code $C$.  Note that 
\begin{equation}
\EE_C \bigl[ 1[d_H(C(t),z) \leq r] \Pr[Z=z | S=s] \bigr]
= \EE_C \bigl[ 1[d_H(C(t),z) \leq r] \bigr] \EE_C\bigl[ \Pr[Z=z | S=s] \bigr],
\end{equation}
since $C(s)$ and $C(t)$ are independent random variables (since $s\neq t$).  We have the following bound:
\begin{equation}
\EE_C[ 1[d_H(C(t),z) \leq r] ] = \Pr_C[d_H(C(t),z) \leq r]
= 2^{-n} \sum_{a=0}^{\lfloor r \rfloor} \tbinom{n}{a}
\leq 2^{-n} 2^{nh(r/n)},
\end{equation}
where we used a tail inequality from \cite[p.39]{welsh} (note that $r \leq n/2$, since $\tau \leq \tfrac{1}{2} - p_e$).  Hence, plugging into (\ref{eq-bat}), we get that 
\begin{equation}
\EE_C \Pr[\Shat \neq S] \leq e^{-2\tau^2 n} + 2^k 2^{-n(1-h(r/n))}.
\end{equation}

Note that $h$ is a concave function, so it satisfies the linear upper-bound $h(r/n) = h(p_e + \tau) \leq h(p_e) + \tau h'(p_e)$, where $h'(p_e) \approx 2.5431$.  Plugging this in, and using equation (\ref{eq-penguin}), we get that 
\begin{equation}
\begin{split}
\EE_C \Pr[\Shat \neq S] &\leq e^{-2\tau^2 n} + 2^k 2^{-n(1-h(p_e)-\tau h'(p_e))} \\
 &= e^{-2\tau^2 n} + 2^{-n\theta} 2^{n\tau h'(p_e)} \\
 &\leq e^{-\Omega(n)}.
\end{split}
\end{equation}
We then use Markov's inequality to get the desired result.  $\square$


\end{document}